\documentclass[preprint2]{aastex}
\usepackage{psfig}
\usepackage{natbib}

\begin{document}
\newcommand{\kms}{km~s$^{-1}$}
\newcommand{\Msun}{M_{\odot}}
\newcommand{\Lsun}{L_{\odot}}
\newcommand{\ML}{M_{\odot}/L_{\odot}}
\newcommand{\etal}{{et al.}\ }
\newcommand{\hhh}{h_{100}}
\newcommand{\hsq}{h_{100}^{-2}}
\newcommand{\tn}{\tablenotemark}
\newcommand{\mdot}{\dot{M}}
\newcommand{\p}{^\prime}
\newcommand{\kmsMpc}{km~s$^{-1}$~Mpc$^{-1}$}

\title{Midlife Crises in Dwarf Galaxies in the NGC 5353/4 Group}

\author{R. Brent Tully,}
\affil{Institute for Astronomy, University of Hawaii, 2680 Woodlawn Drive,
 Honolulu, HI 96822, USA}

\and

\author{Neil Trentham}
\affil{Institute of Astronomy, Cambridge University, Madingley Road, Cambridge CB3 0HA, UK}

\begin{abstract}
This third paper in a series about the dwarf galaxy populations in groups within the Local Supercluster concerns the intermediate mass ($2.1 \cdot 10^{13} \Msun$) NGC~5353/4 Group with a core dominated by S0 systems and a periphery of mostly spiral systems.  Dwarf galaxies are strongly concentrated toward the core.  The mass to light ratio $M/L_R=105~ \ML$ is a factor 3 lower than for the two groups studied earlier in the series.  The properties of the group suggest it is much less dynamically evolved than those two groups of early type galaxies.  By comparison, the NGC~5353/4 Group lacks superluminous systems but has a large fraction of intermediate luminosity galaxies; or equivalently, a luminosity function with a flatter faint end slope.  The luminosity function for the NGC~5353/4 Group should steepen as the intermediate luminosity galaxies merge.   Evidence for the ongoing collapse of the group is provided by the unusually large incidence of star formation activity in small galaxies with early morphological types.  The pattern in the distribution of galaxies with activity suggests a succession of infall events.  Residual gas in dwarfs that enter the group is used up in sputtering events.  The resolution of midlife crises is exhaustion.

\end{abstract}


\section{Introduction}

Galaxy groups and clusters have a wide range of properties.  Membership can vary from a few to thousands of galaxies and those galaxies can vary in type from systems predominantly manifesting the characteristics of young populations to systems that are predominantly old, red, and dead.  This paper is part of a series that explores the nature of groups in a variety of environments.  It was initially motivated by the apparent contradiction between the large numbers of low mass halos anticipated by simulations within the framework of cold dark matter cosmology and the relatively small numbers of dwarf galaxies seen in environments like the Local Group \citep{1999ApJ...522...82K, 1999ApJ...524L..19M}.   Does this evident absence of small systems represent a challenge and constraint on the standard hierarchical clustering paradigm or are small systems being suppressed by astrophysical processes?

Modern large format imagers on large telescopes provide the capability to survey wide areas of the sky to deep levels.  The capability to cover a lot of area makes it feasible to study nearby groups which can be probed to very faint intrinsic luminosities.  Our understanding of the environments of nearby groups is reasonably detailed since redshift surveys provide a rich picture of the surroundings.  As a consequence, the study of groups can be extended to relatively insubstantial structures.  

The program began with a double-blind optical-HI survey of a fraction of the nearby Ursa Major Cluster \citep{2001MNRAS.325..385T}.  
This initial survey established that optical surveys reach deeper than current HI surveys in the sense that all HI candidates were identified in the optical search while only a modest fraction of optical candidates were identified in the HI search.  The HI mass limit in the Ursa Major study was $10^7 \Msun$ and it is probable that many of the optically identified sources would have been picked up with an order of magnitude more sensitivity in the HI observations.  An advantage of an HI detection is access to a redshift which definitively confirms group membership and provides kinematic information.  A clear conclusion of the Ursa Major study was that, in this relatively rich but low density and irregular cluster, the dwarf galaxy population relative to the giant population is small. 
The evidence was already suggesting strong environmental variations in the dwarf/giant dependence and we suggested that the variations may be a signature of the relative epochs of dwarf galaxy formation and reionization \citep{2002ApJ...569..573T}.

To explore the matter further, it was evidently necessary to probe for dwarfs to faint limits over a lot of sky in a wide variety of environments.  The imagers were available at Mauna Kea observatories on the Subaru and Canada-France-Hawaii telescopes.   The plan became to observe the entire virialized regions around groups of a variety of types within the domain of the Local Supercluster, never farther than a distance of 30~Mpc.   In this nearby volume, dwarfs can be detected as faint as $M_R \sim -10$ and
inclusion of all group members except extreme compacts is reasonably assured to $M_R=-12$.  For observational convenience, the program began with investigations of two compact, high density groups with overwhelmingly early-type populations, those centered on the dominant galaxies NGC~5846  \citep{2005AJ....130.1502M} and NGC~1407 \citep{2006MNRAS.369.1375T} (hereafter MTT05 and TTM06).  

The group chosen for the present study has a large mix of spiral galaxies although it turns out in retrospect that  early-type galaxies play a key role.  The most luminous galaxy in the group is the Sb system NGC~5371 with $M_R=-22.3$ at our assumed distance of 29~Mpc.  However it turns out that the dynamical center is coincident with the second and third brightest galaxies, the S0 pair NGC~5353 and NGC~5354.  These two galaxies are separated by only 9~kpc in projection and once they fully merge in the near future their combined luminosity will make the resultant galaxy dominant in the group. 

\section{The Region Surrounding the Survey Area}

At the outset, the type of group we were looking for was one that is reasonably compact yet rich in spirals.  Our Ursa Major case gave us a group that is rich in spirals but is very extended, with no distinct core.  As with the other cases in this series, we looked for a group that is as free as possible from contamination from the near foreground and background so candidate dwarf galaxies would be highly likely group members.   We chose to give attention to a region around the luminous spiral galaxy NGC~5371 which, as we finally constitute the group, contains 12 galaxies of type Sab and later and 10 galaxies of type Sa and earlier with $M_R<-18$.  In the Nearby Galaxies Catalog \citep{1988ngc..book.....T} this group has the numerical designation 42-1.  It is the main knot in a filamentary structure called the Canes Venatici -- Camelopardalis (42) Cloud.  Most compilations of nearby groups contain this structure.  It is group 69 in the catalog of \citet{1982ApJ...257..423H} and \citet{1993ApJS...85....1N} has it as group 124.  It took the referee to point out that it contains Hickson Compact Group 68 as a sub-component \citep{1982ApJ...255..382H}.

\begin{figure}[htbp]
\begin{center}
\includegraphics[scale=0.38]{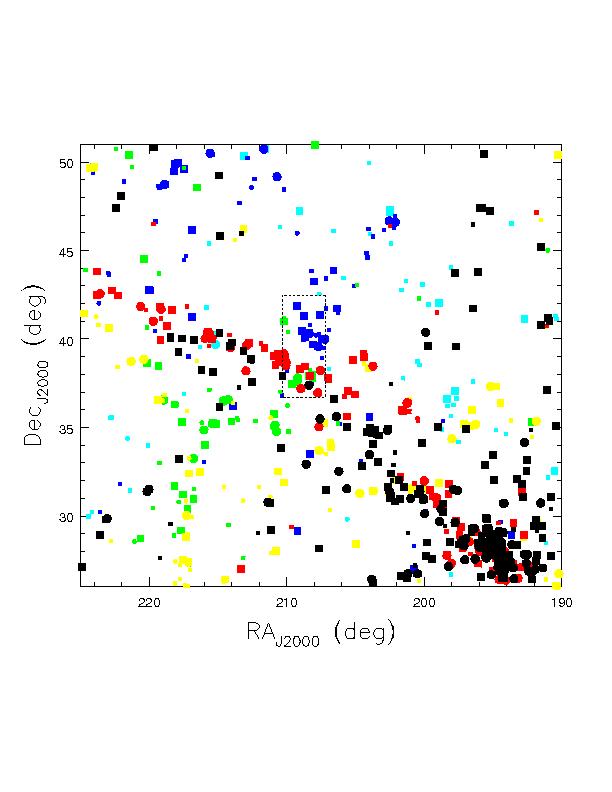}
\caption{Projected positions of galaxies with redshifts $< 8000$~\kms\ in a large region around the NGC~5353/4 Group.  Colors are coded by redshift: cyan $<2000$ \kms, blue 2000--3000 \kms, green 3000--4200 \kms, yellow 4200-5200 \kms, red 5200-6500 \kms, black 6500-8000 \kms.  Large symbols: $M_B<-19.1$ (H$_0=75$); small symbols: $M_B>-19.1$.  Circles: type Sa or earlier; squares: type later than Sa.  The CFHT wide field imaging survey was conducted within the dashed rectangle.  The  Coma Cluster is the dense knot of galaxies in the SW corner. }
\label{filaments}
\end{center}
\end{figure}

A large area around the region of interest is illustrated in Figure \ref{filaments}.   Galaxy redshifts are coded by color and size.  The database used to construct this figure predates the Sloan Digital Sky Survey (SDSS) but we start by showing this version because it is not too busy and it represents what we knew when the study began.  The region that is surveyed in detail is indicated by the dashed rectangle near the center.   The galaxies with heliocentric velocities less than 3000 \kms\ are given large symbols and blue colors.  The 42-1 Group is the nest of blue objects in the upper part of the survey rectangle.  Although there are background structures in the vicinity, and they even pass through the bottom of the survey region, the area of the group itself is clear out to
beyond 8,000~\kms.

\begin{figure}[htbp]
\begin{center}
\includegraphics[scale=0.75]{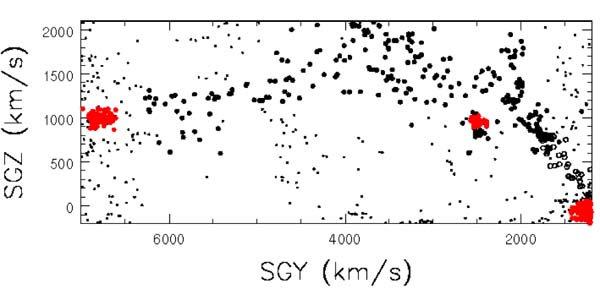}
\caption{Projection in supergalactic SGY-SGZ coordinates with $-300 < {\rm SGX} < 1200$ \kms.  The NGC 5353/4 Group is seen in red at SGY=2500 \kms\ and SGZ=1000 \kms.  The Virgo Cluster is in the lower right corner in red and the Coma Cluster is at the left edge in red.  The heavy filled symbols identify a filament that runs from the Coma Cluster through the NGC~5353/4 Group and meets the plane of the Local Supercluster behind the Virgo Cluster as what is called the Canes Venatici -- Camelopardalis Cloud in the Nearby Galaxies Catalog.  A shortcut to the Virgo Cluster is provided by the minor feature identified by the open circles and identified as the Canes Venatici Spur in the Nearby Galaxies Catalog.  The small dots locate other galaxies seen in projection within this volume. }
\label{3D}
\end{center}
\end{figure}

The large knot of galaxies in the lower right corner of Fig.~\ref{filaments} is the Coma Cluster.  It appears that there is a filament running diagonally toward the upper left from Coma through the bottom of our survey region.  In three-dimensions, it turns out that this feature is the superposition of three structures.  The roughly orthogonal view of Figure~\ref{3D} helps us understand the situation.
One structure runs from Coma forward to the group that we are studying and is highlighted in the figure.  Another is a branch off this filament at 5-6000~\kms\ and runs to high SGZ.  The third is part of the Great Wall at the velocity of the Coma Cluster.  Our 42-1 Group is at a junction involving the filament running in from Coma from the SW, the 42 or CVn--Cam Cloud (in the nomenclature of the Nearby Galaxies Catalog) running in from the N at lower velocities, and the continuation of that structure to the SE at higher velocities.   The other structure of mild concern is a wispy filament (43 or  Canes Venatici Spur in the Nearby Galaxies Catalog) that passes nearby as it links the 42 cloud to the Virgo Cluster which lies at RA=185, Dec=+12.   Galaxies in the survey region with velocities near 1500~\kms\ lie in this structure.  

\begin{figure}[htbp]
\begin{center}
\includegraphics[scale=0.42]{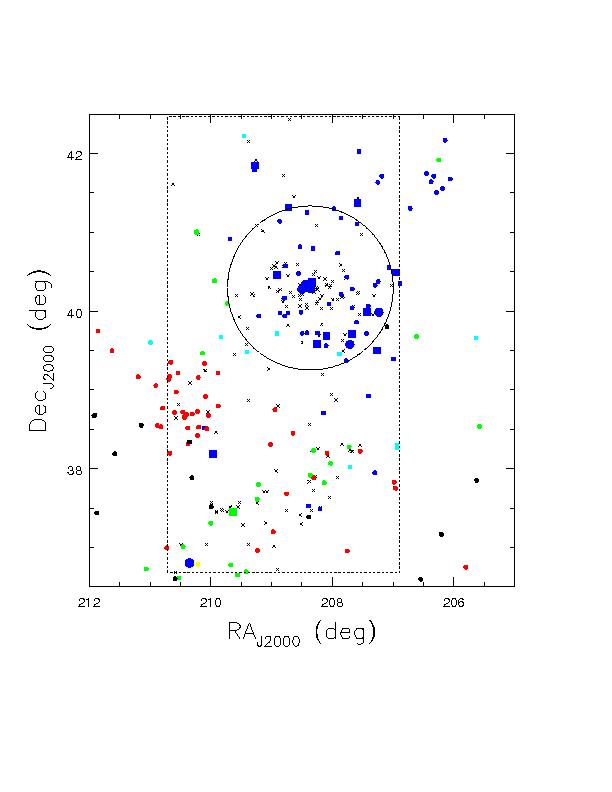}
\caption{CFHT imaging survey area within dotted rectangle.  Galaxies with known velocities are given colors coded by redshift: cyan $<1900$ \kms, blue 1900-3000 \kms, green 3000-4200 \kms, yellow 4200-5200 \kms, red 5200-6500 \kms, black 6500-8000 \kms.  Black crosses identify group member candidates without known velocities.  The more luminous galaxies within the CFHT survey region and in the velocity range of the NGC 5353/4 Group are given larger symbols.  Filled circles: type Sa or earlier, squares: type later than Sa.  The circle identifies the region of the NGC~5353/4 Group.}  
\label{surveyarea}
\end{center}
\end{figure}

\begin{figure}[htbp]
\begin{center}
\includegraphics[scale=0.38]{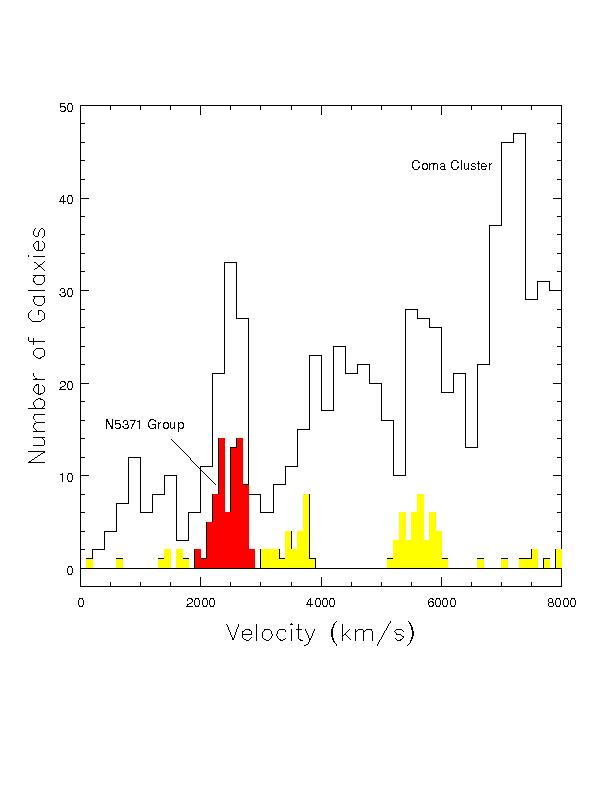}
\caption{Histograms of observed velocities.  Open: velocities in the larger region of Fig.~\ref{filaments}.  Shaded: velocities in the CFHT survey region; highlighted: the velocity range of the NGC 5353/4 Group. }
\label{velocities}
\end{center}
\end{figure}

Figure~\ref{surveyarea} focuses in on the specific area of the wide-field survey described in the next section.   Our deep imaging covers the rectangular area outlined by dotted lines. The colored symbols specify velocities, now reflecting all information available from SDSS data release 5 (data release 6 has subsequently become available but there is little that is new in the NGC~5353/4 region).\footnote{N5353-053 = J135139.3+404414 has a velocity that is correct in DR5 but erroneous in DR6.}  The black crosses identify small galaxies without known velocities that are candidate group members.     Figure~\ref{velocities} shows a histogram of known velocities in the region.  The open histogram relates to the larger region seen in Fig.~\ref{filaments} while the filled histogram relates to the survey region within the dotted rectangle of Fig.~\ref{surveyarea}.  It is seen from this latter figure that there is almost no contamination from objects with known velocity less than 8000~\kms\ in the area of primary interest.  Within the circle to be described later there are two galaxies with velocities below 1800 \kms\ near the S edge.  A couple of galaxies with $V \sim 3500$~\kms\ are just outside the E boundary.  These galaxies will be excluded in the final definition of the group.

It is seen in Fig.~\ref{surveyarea} that
the imaging survey was extended well S of the main clustering of galaxies at $\sim 2400$~\kms.  These observations were made to understand the relationship with galaxies in the adjacent region with slightly larger velocities.  A knot of galaxies with $3000  <  V_h < 4000$~\kms\ is seen in the S third of the survey region and a knot of galaxies with $5000 < V_h < 6000$~\kms\ is seen at the E edge of this region.

\section{The Imaging Survey}

The observations were made with wide-field imagers on the Canada-France-Hawaii Telescope (CFHT). The program was begun in the spring semester of 2002 with the camera called CFH~12k which provided a field of 42x28 arcmin with 0.21" pixels.  It was completed in the spring of 2003 with MegaCam which has a field of 1x1 deg and 0.19" pixels.   A total of 17 sq. deg. were surveyed.  Observations were made in queue mode with sub-arcsecond image quality and photometric conditions.  During the 2002 semester with CFH~12k, 55x10 min exposures were taken in Cousins $R$ band.  During the 2003 semester with MegaCam, 24x10 min exposures were taken through the Gunn $r$ filter.   Exposures were tiled so that every position away from the edges was observed twice and positions that lie in detector gaps in one exposure were picked up in the companion exposure.  The large overlap assures photometric integrity across the entire survey region.  There was a substantial overlap between the areas observed in Cousins $R$ with CFH~12k and in Gunn $r$ with MegaCam which assures a continuity of zero point between observations (results quoted in the Cousins $R$ system).  Color term variations are expected to be below 0.1 mag.

Candidate group members were selected as in the previous papers in this series.  In 17 sq. deg. of sky, over 30,000 galaxies would be expected at our magnitude limit of $R=22$.  Over 99\% are in the background.     We are interested in the low luminosity galaxies which have the distinguishing property that they are  low surface brightness.  \citet{1991A&A...252...27B}  show that dwarf galaxies have fainter central surface brightnesses at fainter absolute magnitudes.  The criteria used in this work are designed to select any galaxies below the upper envelope of the region occupied by dwarf galaxies on the magnitude--central surface brightness plot of Virgo Cluster members in the \citet{1991A&A...252...27B} sample.    Following  \citet{2002MNRAS.335..712T},  galaxies are considered with inner and outer concentration parameters (ICP and OCP)
\begin{equation}
ICP = R(4.4") - R(2.2") < -0.7~{\rm mag}
\end{equation}
\begin{equation}
OCP = R(12") - R(6") < -0.4~{\rm mag}
\end{equation}
where $R(n")$ is the $R$ magnitude within a radius of $n$ arcsec. 
Faint galaxies ($R<20$) are given consideration if they extend across more than 2 arcsec$^2$ with surface brightness $\mu_R < 24.5$~mag~arcsec$^{-2}$.  These selection rules work because dwarfs of all types are found to have light distributions that fall off roughly exponentially with radius \citep{1983ApJ...266L..17F, 1998ARA&A..36..435M}.  The fixed radii that are selected for the indices are appropriate for the distance range 20--30 Mpc which encompasses the range of the groups that have been considered so far in this series.

Our surface brightness criteria accept most dwarfs (with success that is discussed further below) but it admits some background galaxies.  These are discriminated on a basis of morphology.  Background galaxies that pass the surface brightness test are usually distinguished by central high surface brightness components, or hints of spiral structure, or indications of tidal interactions.  We qualitatively rate all the low surface brightness candidates as (1) probable members, (2) possible members, (3) plausible members, and (4) probable background.  Our terminology has turned out to be conservative.
MTT05 and TTM06 provide subsequent evidence demonstrating that galaxies rated 1 and 2 are almost always group members, galaxies rated 3 (the plausible members) are group members over half the time, while galaxies rated 4 are almost always background.  In other words, the only real doubt lies with the candidates rated 3.  These conclusions will be reviewed in the light of the current observations.

The surest information for or against group membership is given by velocities.  The most luminous galaxies had known velocities at the inception of the study.  These initial group candidates were rated `0' ab initio.  In other cases, velocities became available after the candidate search phase of our program.  Those for low surface brightness systems help us evaluate our rating criteria.
If new velocities consistent with group membership are  of galaxies of sufficiently high surface brightness that they were excluded by our ICP/OCP filters the galaxies enter our catalog with a rating of 5.   

Table 1 provides information on 215 candidates within the wide-field imaging survey region.  These candidates either have known velocities consistent with the group or they have morphological ratings 1-3.  Two apparent magnitudes are given for each galaxy. $R_{350}$ is the Cousins $R$ magnitude within a metric aperture of 350 pc radius, described later, and $R$ is a magnitude in a circular aperture at the radius where the galaxy blends into the sky at $\sim 25$~mag arcsec$^{-2}$.   Since extreme dwarf galaxies have surface brightnesses that are very close to the sky limit, measured magnitudes fall short of total magnitudes by amounts that are quite uncertain.  \citet{2006MNRAS.369.1375T} compared observations taken with our current CFHT setup with Subaru Telescope observations that are 1 mag  deeper against the sky.  The Subaru sky limited magnitudes are $\sim 0.2^m$  brighter on average.

\section{ A More Precise Definition of the Group}

We benefit tremendously from the SDSS data release 5 (and subsequently 6) which makes spectra available for most galaxies in our field brighter than $R \sim 17$, $M_R \sim -15$ for group members \citep{2007ApJS..172..634A} .  There are 132 galaxies with $V_h < 8000$~\kms\ in the immediate survey region and 32 more in the larger area shown in Fig.~\ref{surveyarea}.

It is evident that the general definition of the group conforms to the area on the sky within the circle drawn in Fig.~\ref{surveyarea} and to the velocity interval highlighted in the histogram in Fig.~\ref{velocities}.  Galaxies with heliocentric velocities $1900 < V_h < 2900$~\kms\ are considered for membership.  The high velocity limit  of the current sample is secure because there are no galaxies in the vicinity of the circled region with velocities in the interval 2850-3730 and those above 3730~\kms\ fall just outside the circle.  The situation on the low side  is only slightly less well defined.  The 2 low velocity galaxies within but near the S boundary of the circle have velocities 1696 and 1619 \kms.  These galaxies and two more just outside the circle are most plausibly part of the foreground Canes Venatici Spur. 

Luminosity--linewidth distance measurements are available for 4 galaxies within the circled region of Fig.~\ref{surveyarea} \citep{2007arXiv0705.4139T}: NGC 5320, NGC 5346, NGC 5362, and UGC 8726. The averaged distance determination is $29 \pm 3$~Mpc.  The group will be found to have a velocity of 2529~\kms\ in the rest frame of the Local Sheet.  The Hubble ratio 2525/29 = 87~\kmsMpc\ implies a radial peculiar velocity of almost +400~\kms\ if H$_0 = 74$~\kmsMpc. But this curious situation is a story for another paper.

\begin{figure}[htbp]
\begin{center}
\includegraphics[scale=0.38]{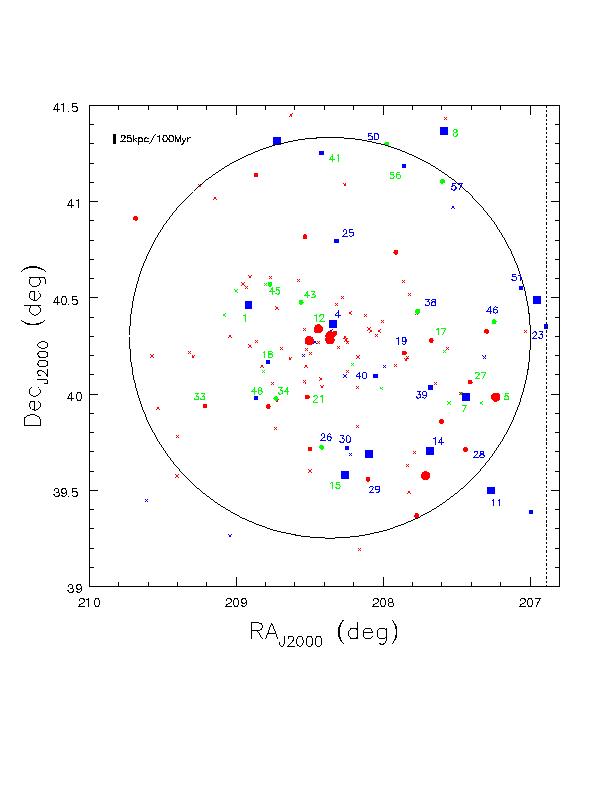}
\caption{The immediate region of the NGC 5353/4 Group.  Only galaxies with $1900 < V_h < 2900$~\kms\ (filled symbols) or membership rating 1--3 (crosses) are included.  Red circles and crosses: types Sa and earlier; blue squares and crosses: Sab and later; green circles and crosses: transition dwarfs typed dE/I.  Circle: centered midway between NGC~5353 and NGC~5354 with radius $1.04^{\circ} = 530$~kpc.   Galaxies identified in Table~2 that show evidence of recent star formation in SDSS spectra are identified by the numbers in the figure.  For those in blue type, the star formation is sufficiently vigorous to drive H$\beta$ into emission.  For those in green type, the star formation is less prominent and H$\beta$ is in absorption.  In the upper left corner of the figure, the vertical bar indicates the projected distance a galaxy will move with the $1 \sigma$ velocity dispersion of the group in 100~Myr.}
\label{types}
\end{center}
\end{figure}

We zoom in with Figure~\ref{types} to the region contained by the circle.  Now, only the galaxies with $1900 < V_h < 2900$~\kms\ or without velocities and with candidate ratings 1-3 are shown.  Galaxies typed earlier and later than Sa are identified by distinct colors and symbols.

\begin{figure}[htbp]
\begin{center}
\includegraphics[scale=0.38]{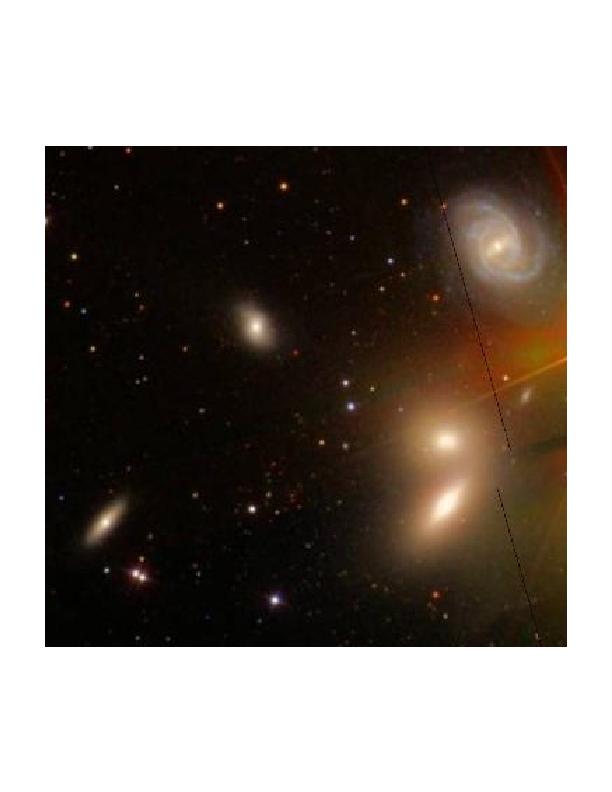}
\caption{Central 85 kpc diameter region of NGC 5353/4 Group in an image from SDSS. NGC 5353 is to the bottom right,
NGC 5354 is immediately above it, NGC 5350 is the barred spiral in the upper right,  and NGC 5355 and NGC 5358 are the other two bright galaxies.  The five bright galaxies in this image constitute HCG~68.}
\label{centralknot}
\end{center}
\end{figure}

It is clear why the circle is centered roughly as it is.  In detail, a further zoom (Figure~\ref{centralknot}) gives a scale for a satisfactory look at the 5 large galaxies near the group center.  The dominant pair that one suspects are in the process of merging are NGC~5353 and NGC~5354.   They are only 9~kpc apart in projection and differ in velocity by 245~\kms.  The most luminous galaxy in the group is the Sb system NGC~5371 at 230~kpc in projection to the ENE (outside Fig.~\ref{centralknot}).  NGC~5353 and NGC~5354 are the second and third most luminous galaxies and the sum of their luminosities exceeds that of NGC~5371.  Because of their central location and combined luminosity we give the group their name.  The circle shown in the various figures is centered halfway between them.

The five major galaxies seen in Fig.~\ref{centralknot} constitute HCG~68 \citep{1982ApJ...255..382H}.  In this instance the Hickson Compact Group is substructure within a larger bound entity.  There is the high likelihood that the close proximity to the core of the spiral NGC~5350 is a chance projection effect (though there is little doubt that it is a group member).  If this galaxy is ignored, the remainder probably would not have made the HCG list.  Nevertheless, the inclusion in the HCG catalog does draw attention to the unusual nature of the core of the NGC~5353/4 Group.

\begin{figure}[htbp]
\begin{center}
\includegraphics[scale=0.38]{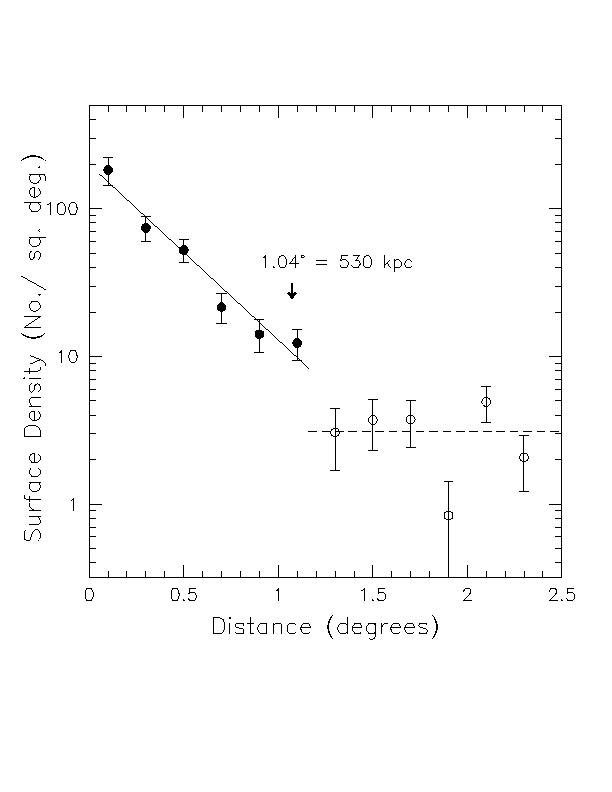}
\caption{Surface density of candidate group members as a function of distance from a point mid-way between NGC~5353 and NGC~5354.  Solid line has a slope of 1.2.  Dashed line defines background level.}
\label{radgrad}
\end{center}
\end{figure}

The surface density of candidate galaxies with respect to this center is shown in Figure~\ref{radgrad}.   It is seen that the surface density falls off steeply as $\Sigma \propto r^{-1.2}$ (corresponding to a volume dependence of $\rho \propto r^{-2.2}$), with an apparent drop at $\sim 1.2^{\circ}$ to a field value of $\sim 3$ galaxies / sq. deg.

There are 53 galaxies with accepted velocities within $1.23^{\circ}$  of NGC~5353/4.  The velocity dispersion of these is $\sigma_V = 205$~\kms\ around a group mean of $V_h=2434 \pm 28$~\kms.  The mass within this region can be calculated alternatively using the virial theorem or the probable mass estimator of  \citet{1985ApJ...298....8H}.   The two methods give masses of $2.19 \times 10^{13} \Msun$ and $2.05 \times 10^{13} \Msun$ respectively and we accept an average of $2.1 \times 10^{13} \Msun$.  The luminosity associated with all 137 candidates within the same radius is $L_R = 2.0 \times 10^{11} \Lsun$ so the mass to light ratio is $105~ \ML$.

MTT05 discuss a parameter, the radius of second turn-around today in a spherical collapse, $r_{2t}$.  It is determined by TTM06 through two alternative expressions
\begin{equation}
r_{2t} = 0.193 (M_{12})^{1/3}~ {\rm Mpc}
\end{equation}
\begin{equation}
r_{2t} = \sigma_V / 390~ {\rm Mpc}
\end{equation}
There is evidence \citep{2006EAS....20..191T} that these relations are valid on mass scales from $10^{12}$ to $10^{15} \Msun$.  In the present case, with mass in units of $10^{12}  \Msun$ of $M_{12} = 21$ and $\sigma_V = 205$~\kms, estimates for $r_{2t}$ are 532 and 526 kpc, respectively.  We accept a value of $r_{2t} = 530 \pm 20~{\rm kpc} = 1.04^{\circ} \pm 0.04$, consistent with the surface density edge for the group seen in Fig.~\ref{radgrad}.  In the following discussion, the group will be considered to include the 137 candidates within $1.23^{\circ} = 625$~kpc of the center identified in Table~1.

What do we learn from the survey coverage beyond the $r_{2t}$ circle?
Within the CFHT survey area at Dec~$< 39^{\circ}$, it is seen in Fig.~\ref{surveyarea} that there are two major concentrations of galaxies: at RA~$\sim 209$, Dec~$\sim 37.5$, $V_h \sim 3500$~\kms\ and at RA~$\sim 210.5$, Dec~$\sim 38.5$, $V_h \sim 5500$~\kms.  There are a modest number of dwarf candidates projected on the concentration at 3500~\kms\ but very few near the concentration at 5500~\kms.  Our procedure for identifying nearby dwarfs evidently still picks up candidates at 3500~\kms\ but not at 5500~\kms.

Only $\sim 25\%$ of the CFHT survey area lies within the circle describing the second turnaround radius for the group.  The large area outside the circle provides a control of our procedures to identify dwarfs.  Candidates are spatially concentrated.  The paucity of scattered candidates indicates that there is not a serious contamination problem.   Most of the candidates outside the circle lie in the vicinity of the structure at 3500~\kms\ to the S.   We will not give further attention to that more distant filament  because the census of dwarfs there is unlikely to be complete.

\section{The Morphological Distribution of Galaxies within the Group}

\begin{figure}[htbp]
\begin{center}
\includegraphics[scale=0.38]{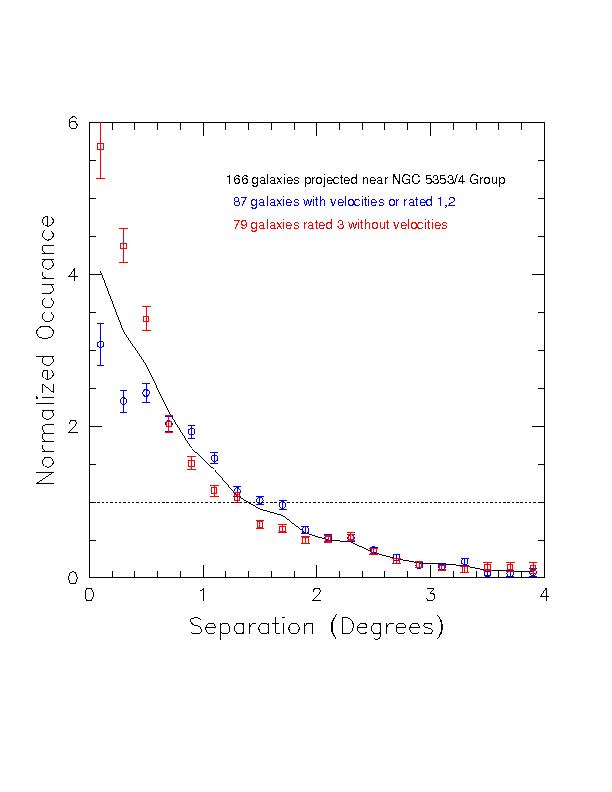}
\caption{Relative 2-point correlation functions.  Black line: all group candidates.  Blue circles: high confidence members (known velocities or rated 1-2).  Red squares: Suspected members (no velocities and rated 3). }
\label{correlations}
\end{center}
\end{figure}

It is evident from Fig.~\ref{types} that galaxies are strongly clustered around NGC~5353/4.  That tendency is made clear in the two-point correlation functions of Figure~\ref{correlations}.  Normalization for these distributions is given by random populations of the survey area with equal numbers of galaxies.  The solid line shows the correlation with all candidates in the survey region with declination greater than $38.5^{\circ}$. The open blue symbols indicate the correlation with the roughly half of these galaxies either with velocities or rated 1 and 2; i.e., galaxies that are almost certainly group members.  The open red symbols give the correlation for the other roughly half the sample that do not have velocities and are rated 3; i.e., less certain group members.  Remarkably, the correlation is stronger with the supposedly less certain members.

This correlation analysis only confirms what is apparent to the eye upon looking at Fig.~\ref{types}.  The galaxies with velocities are the brighter ones and many of these are later types and reside in the outer parts of the group.  The galaxies with less certain membership ratings are faint and usually classed dE.  Their very strong correlation is convincing evidence that almost all of them belong to the group. 

\begin{figure}[htbp]
\begin{center}
\includegraphics[scale=0.37]{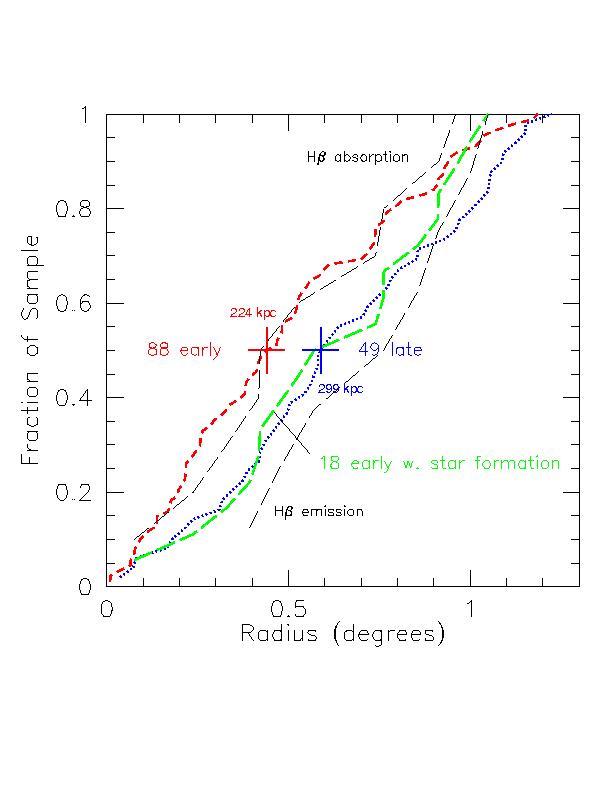}
\caption{Radial cumulative fractions for 137 group candidates.  Dashed red curve: Sa and earlier; 50\% within 224 kpc of center.  Dotted blue curve: Sab and later; 50\% within 299 kpc of center. Long dashed green curve: 18 galaxies with early morphological types with SDSS spectra that manifest recent star formation.  Long dashed black curve on the right side: 8 early morphological types with H$\beta$ in emission.  Long dashed black curve on the left side: 10 early morphological types with H$\beta$ in absorption but spectral indications of star formation. }
\label{cumulative}
\end{center}
\end{figure}

Another way to see the difference between the distributions of early and late galaxies is with the radial cumulative functions of Figure~\ref{cumulative}.  The 88 galaxies within $1.23^{\circ}$ radius typed Sa and earlier are distributed with 50\% of them within 224~kpc of the center ($0.42 r_{2t}$) while the 49 galaxies typed later than Sa are distributed with a 50 percentile radius of 299~kpc ($0.56 r_{2t}$). A Kolmogorov-Smirnov (K-S) test gives the probability that the early and late type radial dependencies are drawn from the same distribution is 4\%.  The K-S test is suggestive but not conclusive evidence that early types are more centrally concentrated than late types.

\section{An Unusual Population of Central Starburst Dwarfs} 

\begin{figure}[htbp]
\begin{center}
\includegraphics[scale=0.48]{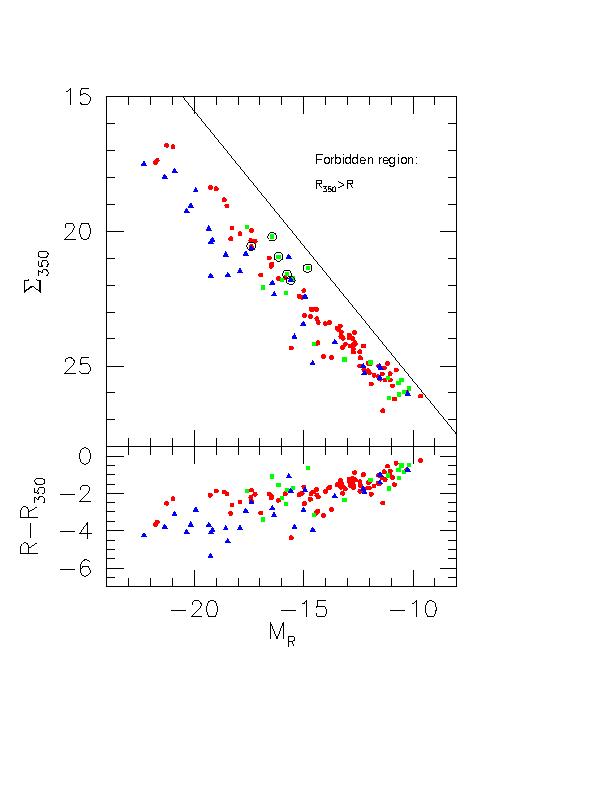}
\caption{Top panel: Mean surface brightness within 350 pc of the galaxy center as a function of total intrinsic luminosity.  Red circles: Sa and earlier.  Blue triangles: Sab and later. Green squares: transition dE/I or ambiguous types. Group memberships are confirmed by redshifts for almost all galaxies brighter than $M_R=-15$.  In 6 cases identified by open circles the galaxies were not suspected to be members in advance of the availability of spectra.  Bottom panel:  The magnitude of the entire galaxy minus the magnitude contained within the central 350~pc aperture as a function of absolute magnitude.}
\label{R300}
\end{center}
\end{figure}

Figure~\ref{R300} is a variation on scaling relations similar to those introduced by MTT05 and TTM06.  $R_{350}$, the $R$ band magnitude within the metric radius of 350 ~pc, $\ell_{350} = 2.5$~arcsec, is a measure of the central surface brightness of the group members.  The mean surface brightness within this metric radius is $\Sigma^R_{350} = R_{350} + 2.5 {\rm log} (\pi \ell_{350}^2)$.  There is a pronounced correlation between central surface brightness and absolute magnitude.  Early and late types are given different symbols.  Late types tend to have fainter central surface brightnesses at a given total magnitude, $M_R$.  

A third, intermediate class of galaxy is also identified in Fig.~\ref{R300}.  These are galaxies classified
dE/I.  Galaxies have received this classification for two different reasons.  In cases with $M_R > -13$ this classification is given because of ambiguity in instances where there is not much information because the candidates are so small.  In the other cases, which are the ones that really interests us, the ten candidates lie in the range $-18 < M_R < -14$.  These galaxies would be classified dE on morphological grounds but the SDSS spectra have emission line features and, upon close inspection, high surface brightness lumpy structure is usually seen at the galaxy centers.  We refer to these galaxies as transition cases.  Similar objects have been seen in the Virgo Cluster  \citep{1991A&A...252...27B, 2006astro.ph..8294L} and in the field \citep{2006AJ....131..806G}.

The SDSS spectra provide very valuable information.  Of 53 group members with velocities, 42 have spectra accessible through the SDSS archive.  Table~2 provides a summary of the properties of those spectra.  Galaxies are ordered in this table from those with spectra revealing the most vigorous ongoing star formation to those with spectra of old Population II.  The first group of 13 galaxies have spectra with
the Balmer series from H$\alpha$ to H$\delta$ in emission.  Progressing down the table, first H$\delta$ passes into absorption then the other Balmer lines follow in turn.  After H$\alpha$ has transitioned into absorption there are still a few cases with weak emission in [SII] and [NII].  Three of these have strong H$\beta$ through H$\delta$ absorption indicative of a post-starburst A star contribution.  Finally, the last 8 cases with SDSS spectra are characteristic of Population II.

The spectra suggestive of ongoing star formation also contain the familiar diagnostics of metallicity.
High luminosity systems have strong [NII] and weak [OIII] indicative of high metallicity and low luminosity systems have weak [NII] and strong [OIII] as found in low metallicity systems.

We find the high fraction of galaxies with the features of ongoing or recent star formation to be remarkable, particularly because the incidences are so frequent in galaxies that have little morphological texture.   Consider some statistics.  We ignore the brightest galaxies because only 9 of 15 with $M_R<-19$ have SDSS spectra.  By contrast, 33 of 38 galaxies with known velocities at $M_R>-19$ have SDSS spectra.   Among these 33, 24 are typed Sa or earlier, including the transition dE/I galaxies in this category.  Surprisingly, 18 of these 24 have signatures of recent star formation.  It can be seen by scanning Table~2 that there is little correlation with morphology in the sequence from vigorous star formation to weak manifestations of recent star formation.   The only non-surprise is that the 8 cases with pure Pop II spectra are all typed early.      

It is interesting to see where the objects classed morphologically early but with spectral signatures of recent star formation are located in the group.  The radial cumulative plot, Fig.~\ref{cumulative}, shows that these objects have a radial distribution more like the late than the early types.  A K-S test that compares the radial distributions of the 18 early types manifesting recent star formation with the remaining 60 early types with either Pop II spectra or no available spectra gives a probability of 6\% that they are drawn from the same distribution.  The relative lack of central concentration of the systems with recent star formation suggests that they are not a dynamically relaxed population within the group.
Further, if the sample of 18 early type galaxies with evidence of star formation in SDSS spectra are split between the 8 with H$\beta$ in emission (most vigorous or recent star formation) and the 10 with H$\beta$ in absorption (less vigorous or recent star formation) those with more pronounced star formation features are less concentrated with respect to the center of the group.

These morphologically early type objects with ongoing or recent star formation are intriguing because we do {\it not} find a substantial population of such objects in either the NGC~5846 or NGC~1407 group studied earlier (MTT05, TTM06).  Those groups are thought to be at a more advanced stage of dynamical evolution.  The most direct comparison can be made with the NGC 5846 Group because it too is contained within the SDSS.  In that case, we find that 17 of 48 SDSS spectra of early morphology types have some evidence of ongoing or recent star formation.  In addition, we have Keck Telescope spectra of 17 relatively faint early morphological types in this group and 5 show manifestations of star formation.  Hence 1/3 of early types in the NGC 5846 Group (which consists of 80\% early types) have evidence of recent star formation compared with 2/3 of early types in the NGC 5353/4 Group.  

\begin{figure}[htbp]
\begin{center}
\includegraphics[scale=0.48]{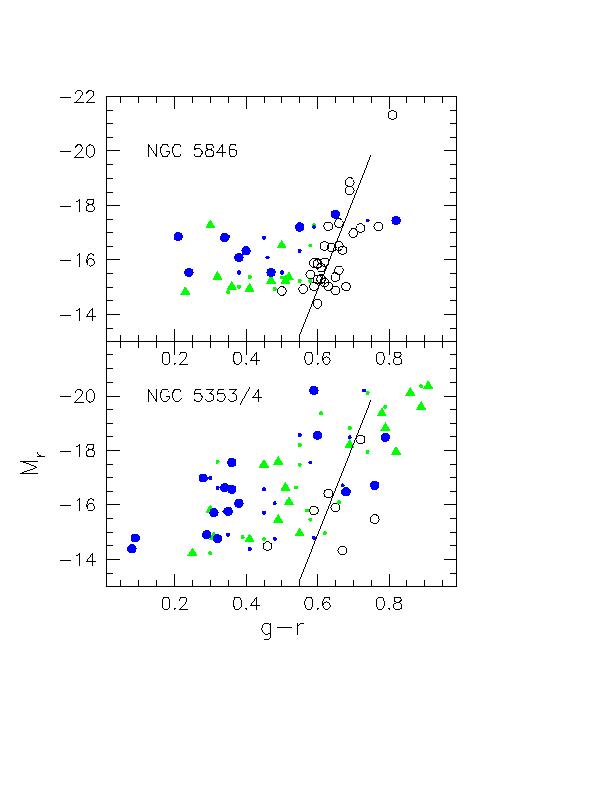}
\caption{Color--magnitude plots for galaxies with SDSS spectra.  Data for the NGC 5353/4 Group is shown in the bottom panel and, for comparison, NGC~5846 Group data is shown in the top panel.  Cases with no sign of recent star formation are indicated by black open circles.  Cases with modest and vigorous star formation are indicated by green triangles and blue circles, respectively.  In instances with recent star formation, large symbols identify the color in the SDSS fiber that extracted the spectrum and the small
symbols identify the global color.  The straight line is a fit to the open circles in the top panel.}
\label{cmd}
\end{center}
\end{figure}

The comparison with the NGC~5846 Group continues over the next four figures. An $r$, $g-r$ color--magnitude diagram is seen in the top panel of Figure~\ref{cmd} for all galaxies in the NCG~5846 sample with SDSS spectra.  Galaxies with no indication of recent star formation are identified by black open circles.  The solid line is a least squares fit to the data for these galaxies with uncertainties taken in color.  The trend toward bluer colors at fainter magnitudes can be understood as the consequence of diminished metallicities.  The filled symbols that scatter to bluer colors represent galaxies with manifestations of star formation in their SDSS spectra.  Blue circles are reserved for cases with both H$\alpha$ and H$\beta$ in emission.  Green triangles denote cases with sufficiently less current star formation that H$\beta$ is in absorption.  Spectroscopic redshifts are available for almost all galaxies in this group brighter than $M_r = -15$ but commonly for cases brighter than $-19$ there is no SDSS spectrum available.

\begin{figure}[htbp]
\begin{center}
\includegraphics[scale=0.48]{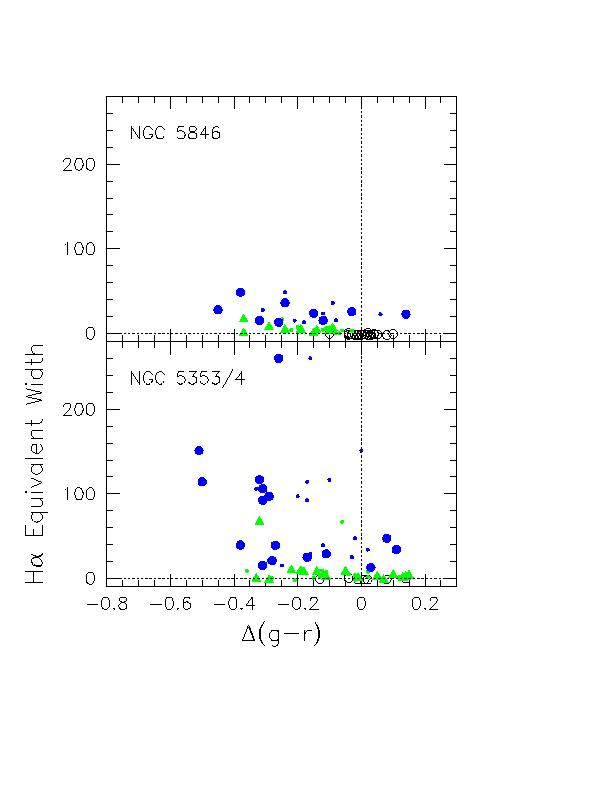}
\caption{Color displacement as a function of H$\alpha$ emission line equivalent width.  Color displacements are with respect to the sloped line in Fig.~\ref{cmd}.  Equivalent widths are extracted from the SDSS database.  Symbols have the same meaning as in Fig.~\ref{cmd}.}
\label{dc-ew}
\end{center}
\end{figure}

\begin{figure}[htbp]
\begin{center}
\includegraphics[scale=0.38]{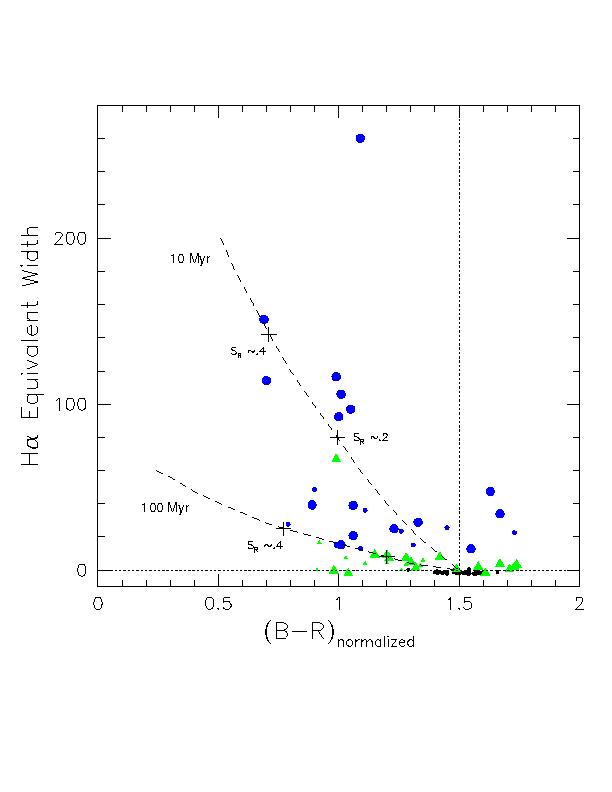}
\caption{Color displacement as a function of H$\alpha$ equivalent width, now transformed from SDSS $g-r$ to Johnson-Cousins $B-R$ and given a zero point such that systems with no hint of star formation (indicated by small black circles) scatter around $B-R=1.5$.  Colors have the same meaning as in the previous 2 figures.  Now, cases drawn from the NGC~5353/4 Group are given big symbols and cases drawn from the NGC~5846 Group are given small symbols.  The dashed curves are carried over from \citet{2003ApJ...582..668B} for models with steep IMF and star formation extending over 10~Myr and 100~Myr respectively.  Crosses on each dashed curve indicate locations where fractions 0.2 and 0.4
of the $R$ band light are produced by the current star formation episode.}  
\label{dBR-ew}
\end{center}
\end{figure}

The equivalent data for the NGC~5353/4 Group is shown in the lower panel.  The quality of the data is similar since the two groups are at similar distances and equally well covered by SDSS.  The sloped line is transposed from the top panel.  It is evident that  the fraction of open symbols identifying systems without recent star formation is much smaller than in the top panel.

The same samples are carried over to Figure~\ref{dc-ew}.  Now, the vertical axis is the equivalent width of H$\alpha$ given by the SDSS database and the horizontal axis is the displacement in color from the sloped line in Fig.~\ref{cmd}.  Clearly, equivalent widths reach much higher values in the spectra associated with the NGC~5353/4 Group.

In Figure~\ref{dBR-ew}, data for the two groups are combined, with those from the NGC~5353/4 Group given large symbols and those from the NGC~5846 Group given small symbols.  The cases with no sign of star formation are reduced to small dots.  Equivalent widths are as in Fig.~\ref{dc-ew} but now colors are transformed \citep{2006A&A...460..339J} to $B-R$ and given zero points consistent with $B-R=1.5$ if no recent star formation.   This color zero point facilitates a comparison with the preferred model among those developed by \citet{2003ApJ...582..668B}.  In this case, star formation is continuous over specified intervals with an initial mass function steeper than Salpeter.  The dashed curves in the plot give loci of varying `strengths' of the episodes over periods of 10~Myr and 100~Myr.  The `strength', $S_r$, is measured by the contribution to the $R$ band light generated by the current star formation activity.  To first order, the blue color shift from $B-R=1.5$ is a measure of the contribution from the current episode of activity, and is found to produce up to 40\% of the observed red light in the cases under consideration.  The equivalent width is a measure of the age of the star formation event.  It is only possible to have large equivalent widths if hot stars have been formed recently.  Characteristic ages of $\sim 10$~Myr are found in quite a few galaxies in the NGC~5353/4 Group as compared with $\sim 100$~Myr for the most active galaxies in the NGC~5846 Group.

\begin{figure}[htbp]
\begin{center}
\includegraphics[scale=0.38]{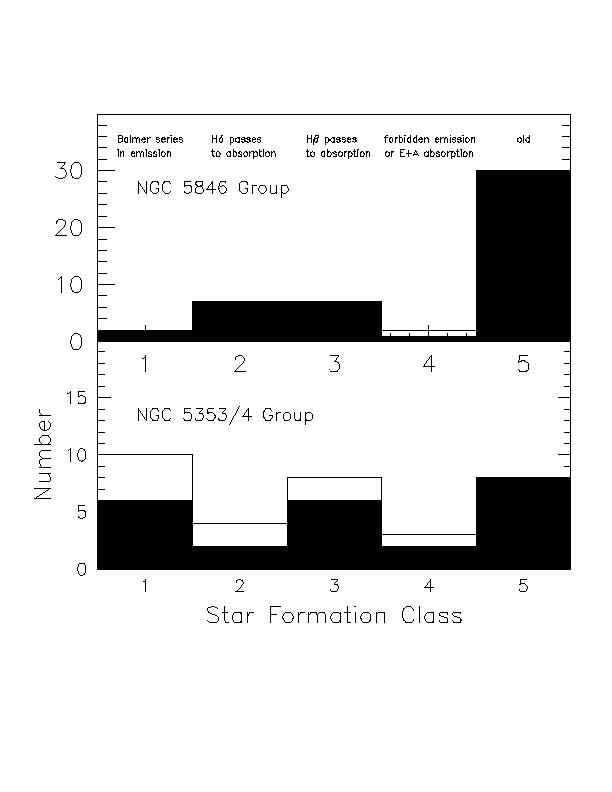}
\caption{Distribution into star formation classes for galaxies with SDSS spectra fainter than $M_r=-19$ in the two groups associated with NGC~5353/4 and NGC~5846.  Filled histogram: types earlier than Sa; open histogram: all types. From left to right: progression from vigorous star formation to no sign of current star formation.}
\label{starform}
\end{center}
\end{figure}

Figure~\ref{starform} provides a summary.  The filled histograms only include data from early type galaxies.  In the case of the NGC~5846 Group, only 1/3 of early galaxies show indications of recent star formation while in the NGC~5353/4 Group only 1/3 of galaxies that appear morphologically textureless lack signatures of star formation.

It was noted in the discussion around Fig.~\ref{cumulative} that early-type galaxies in the NGC~5353/4 Group manifesting a strong star formation episode are typically farther from the center of the group than those showing weaker star formation features.  We can see where the various galaxies are in Fig.~\ref{types}.  The galaxies with stronger and weaker star formation properties are curiously separated.  Those with strong features all lie in the western half of the group and congregate in the south--west quadrant.  Those with weaker features tend to lie to the east and along the mid-plane.  In the upper left corner of the plot is a bar that indicates how far a galaxy will typically move in the group in 100~Myr.  It is inferred that the current episodes have begun essentially where we see the galaxies today.

\begin{figure}[htbp]
\begin{center}
\includegraphics[scale=0.31]{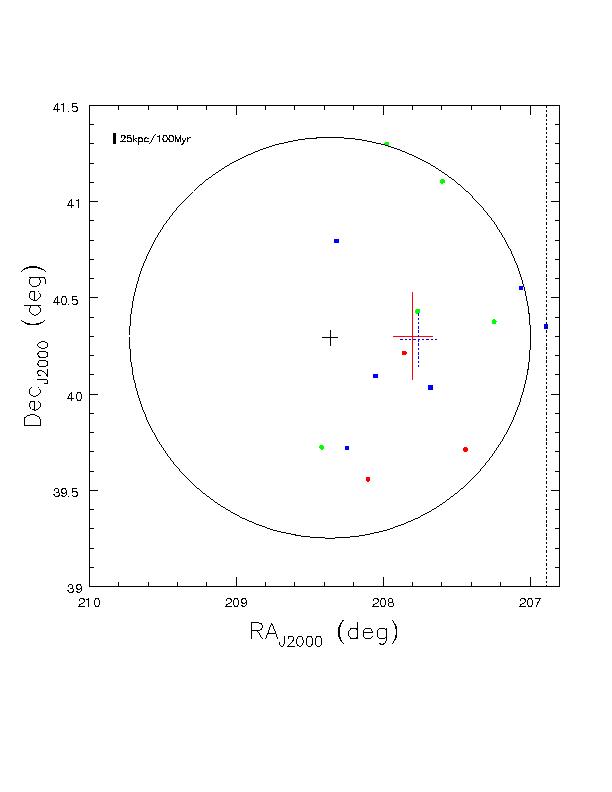}
\includegraphics[scale=0.31]{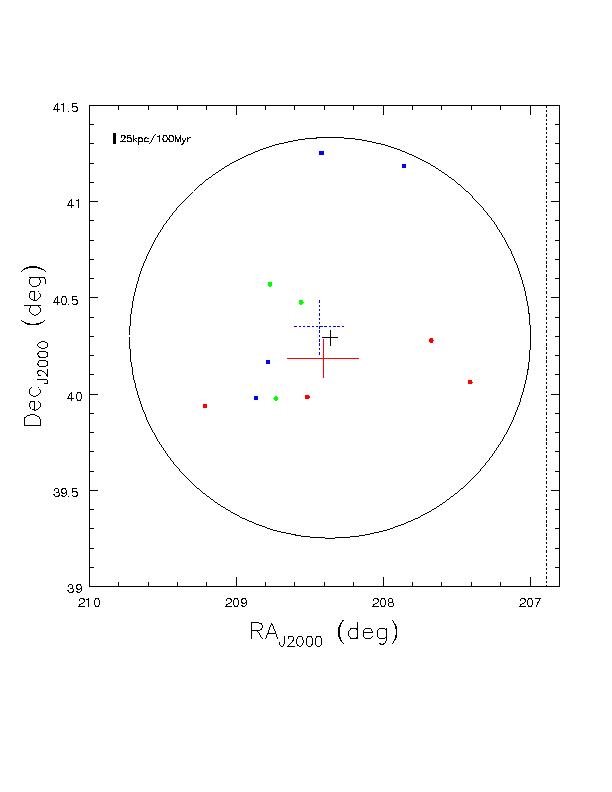}
\caption{{\it Top:} Positions of galaxies with evidence of vigorous current star formation. {\it Bottom:} Positions of galaxies with evidence for an older or weaker star formation episodes.  Red circles: morphologically early types; green circles: transition types that would be called early except for slight lumpiness; blue squares: morphologically late types.  Large solid red crosses: mean positions of the early and transition types, with the lengths of the cross arms equal to $1 \sigma$ deviations.  Dashed blue crosses: mean positions and $1 \sigma$ deviations of all types.}
\label{star1-2}
\end{center}
\end{figure}

Figure~\ref{star1-2} represents an attempt to clarify the significance of the apparent shift in locations of the distinctive populations.   The locations of the 25 galaxies with $M_R>-19$ with manifestations of recent star formation are shown; the most active cases in the top panel and the less active cases at the bottom.  The mean positions and rms deviations are indicated for the separate samples.  The differences in RA have a significance of $3 \sigma$.  The K-S test gives a probability of 3\% that the early and transitional types are drawn from the same RA distribution in the two cases; the probability is 0.3\% if all types are considered.  This 2-dimensional comparison provides stronger evidence than the radial distribution K-S tests for spatial differences between galaxies with distinctive star formation histories.

\begin{figure}
\includegraphics[scale=1.0]{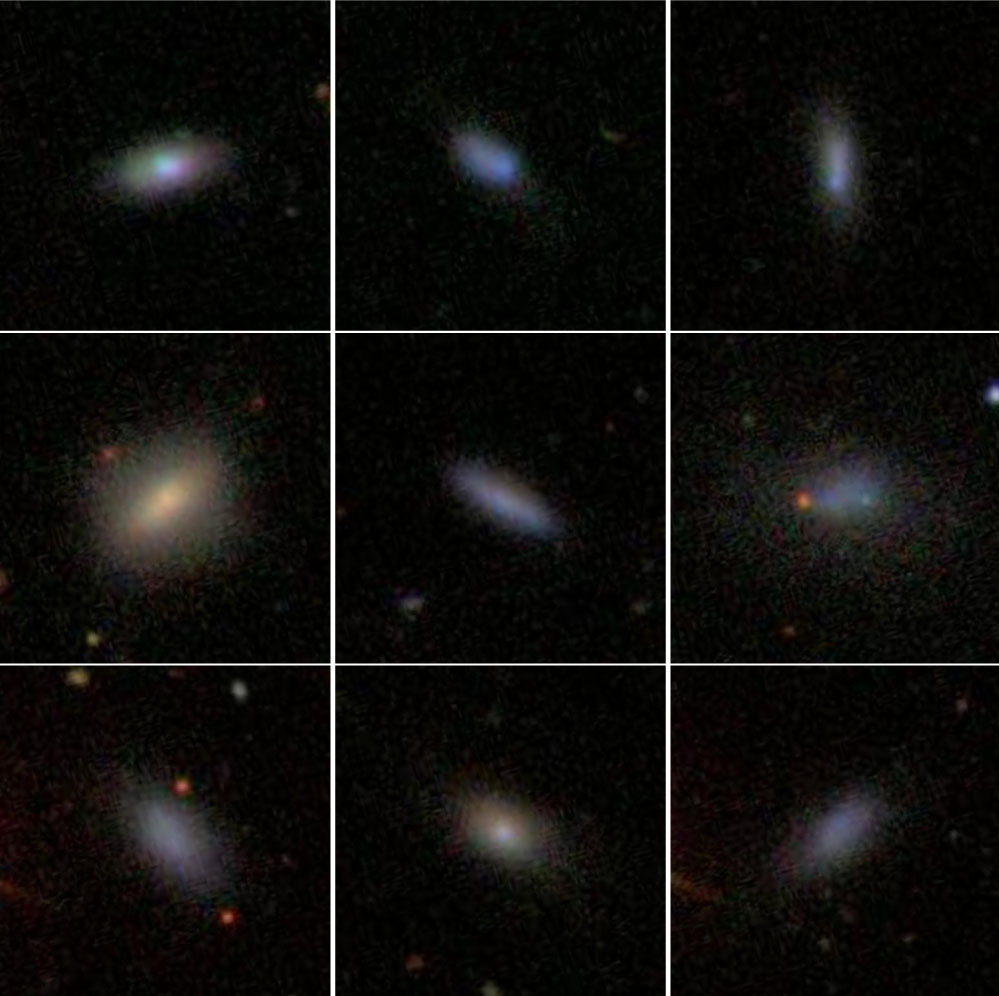}
\caption{SDSS images of 9 early and transition dE/I low luminosity galaxies in the NGC 5353/4 Group with spectral evidence of recent star formation. From left to right and top to bottom: N5353-038, -050, -046, -029, -058, -034, -043, -041, -045.}
\label{images}
\end{figure}

\begin{figure}
\includegraphics[scale=0.74]{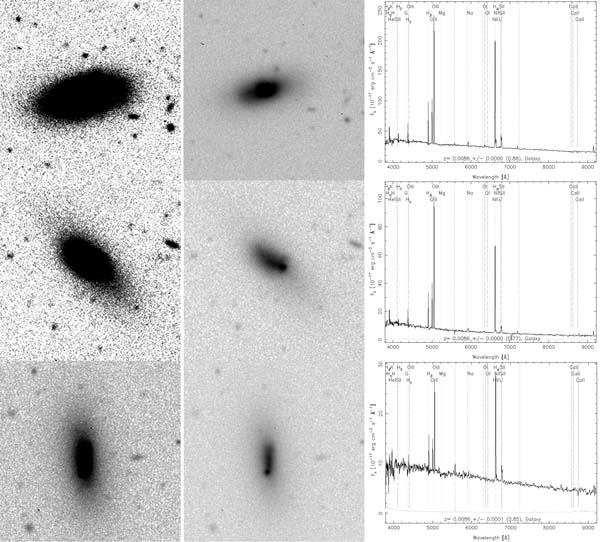}
\caption{Early and transition dE/I low luminosity galaxies in the NGC 5353/4 Group with spectral evidence of recent star formation. $R$ band CFHT MegaCam images are stretched to show outer and central structures, respectively.  Spectra are drawn from the SDSS database. Three cases with vigorous star formation giving $H\beta$ in emission: from top to bottom: N5353-038, -050, -046}
\label{spec1}
\end{figure}

\begin{figure}
\includegraphics[scale=0.74]{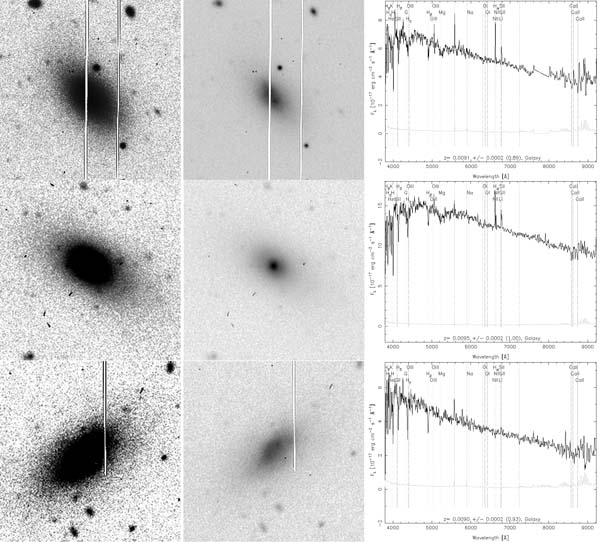}
\caption{Early and transition dE/I low luminosity galaxies in the NGC 5353/4 Group with $H\beta$ in absorption, indicating less recent or intensive star formation. From top to bottom: N5353-043, -041, -045.}
\label{spec2}
\end{figure}

The evidence of activity in galaxies that show little structure is particularly noteworthy.  In the images of 9 examples in Figure~\ref{images}, ordered from vigorous current star formation in the upper left to weak current star formation in the lower right, there are only hints of the activity.  Six of these cases, the 3 with the stronger current star formation in the top row of Fig.~\ref{images} and the 3 with the weaker current star formation in the bottom row of this figure, are given closer attention in Figures~\ref{spec1} and \ref{spec2}.  The pairs of images are based on the CFHT observations.  In the images on the left that emphasize the outer structures each of the galaxies appears ellipsoidal and structureless, the morphological features of dE types.  In the complementary images, the less deep prints bring out the central high surface brightness features, also visible in the images of Fig.~\ref{images}.  The SDSS spectra are given in the right panel.

\onecolumn

\twocolumn

It is normal to observe ongoing star formation in late type systems but exceptional to find it in early type dwarfs.  Some 15\% of brighter dE in the Virgo Cluster evidently are transition galaxies of a similar nature  \citep{2006astro.ph..8294L}.   Many years ago, during a survey in HI of dwarf galaxies \citep{1981ApJS...47..139F}, a related phenomenon was seen in Virgo galaxies.  In the field one could anticipate if a galaxy would be detectable in HI based on its morphology,  but in the Virgo Cluster there are textureless systems that unexpectedly have significant HI.  \citet{2001ApJ...559..791C} have pointed out that the dE population in Virgo has a larger velocity dispersion than the giant ellipticals, suggesting that the two populations are not coeval and the dE must be remnants of later arrivals.
The situation in the NGC~5353/4 Group deserves attention because of the high fraction of this class of object at intermediate luminosities ($M_R \sim -16$) showing spectral evidence of gas reservoirs.  The situation remains to be explored at fainter luminosities;  at $M_R > -15$ there is very little spectral information yet for the NGC 5353/4 Group sample.  In the last section we will speculate on the nature of these patterns of star formation and the possibility that a significant fraction of dwarfs arrive late. 

\section{The Luminosity Function}

\begin{figure}[htbp]
\begin{center}
\includegraphics[scale=0.38]{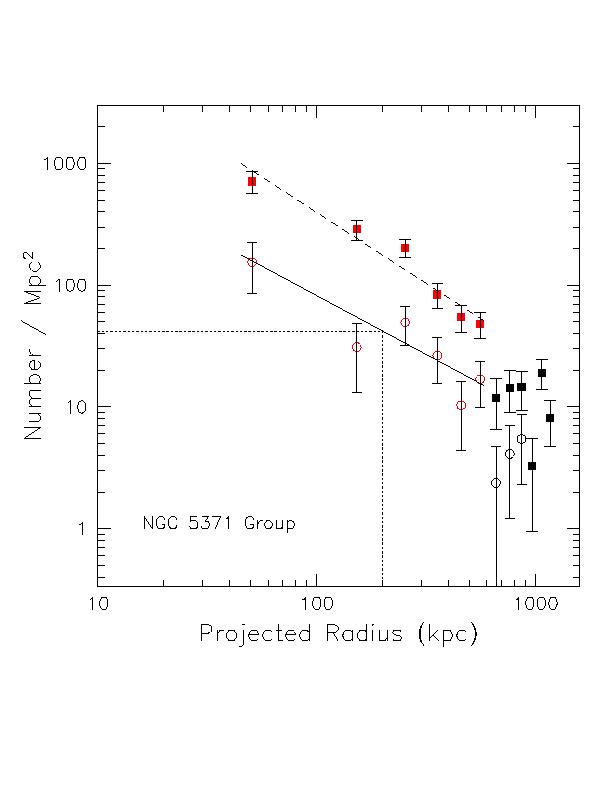}
\caption{Group density normalization.  Filled squares: radial density fall-off with respect to the NGC 5353 - NGC 5354 mid-point for all candidates.  Open circles:  same but only including bright galaxies with $M_R < -17$.  The density of bright galaxies at 200 kpc radius of 42 galaxies Mpc$^{-2}$ provides the luminosity function normalization.}
\label{lglgden}
\end{center}
\end{figure}

We begin this discussion with the group density normalization introduced by \citet{2002MNRAS.335..712T} because the ultimate goal of the program is to intercompare different environments.  Figure~\ref{lglgden} serves this purpose.  The filled squares show the run of surface density for all group candidates while the open circles show the same thing but restricted to galaxies with $M_R < -17$.  Our standardized density measure is the number surface density of luminous galaxies ($M_R < -17$) at a radius of 200~kpc.  In this case, that value is 42 galaxies Mpc$^{-2}$, $\sim 30\%$ higher than in the two groups studied earlier.

\begin{figure}[htbp]
\begin{center}
\includegraphics[scale=0.38]{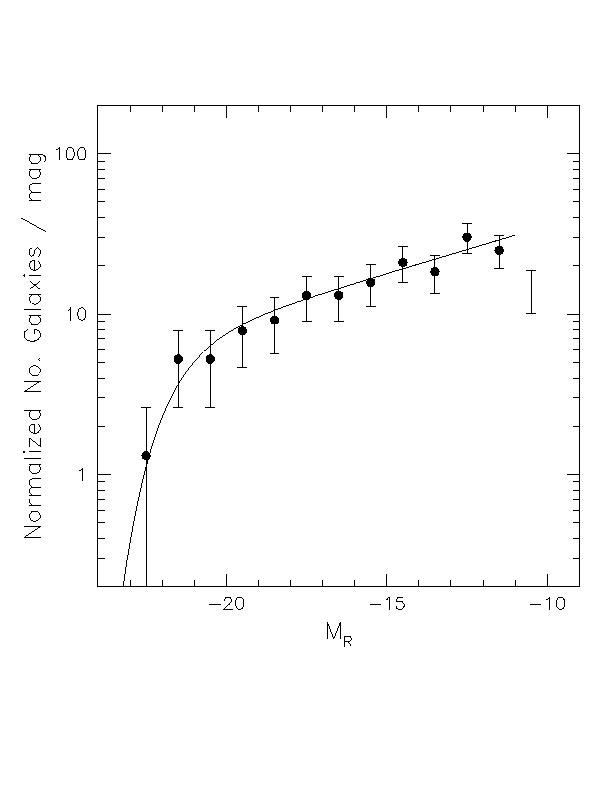}
\caption{Luminosity function for the NGC 5353/4 Group.  All galaxies within $1.23^{\circ}=625$~kpc of the mid-point between NGC 5353 and NGC 5354 with an acceptable velocity or rating 1-3 is included.
Incompletion is taken to set in faintward of $M_R = -11$.  The solid curve is a Schechter function fit described by the parameters $M_R^{\ast}=-21.9$ and $\alpha = -1.15$.  The count per magnitude bin is normalized to sum to 42 for $M_R < -17$.}
\label{LF}
\end{center}
\end{figure}

With this normalization, we plot the NGC~5353/4 Group luminosity function in Figure~\ref{LF}.  We accept all candidates with known velocities in the window $1900 < V_h < 2900$~\kms\ plus all those with membership ratings $1 - 3$.  In MTT05 and TTM06 the luminosity functions for the NGC~5846 and NGC~1407 groups were built with a range of assumptions about the rating 3 candidates but the exceptionally strong  two-point correlation of these candidates in the present sample persuades us to consider all these objects as members.  It is to be noted that almost all the galaxies with $M_R < -15$ are confirmed members on the basis of redshifts.

\begin{figure}[htbp]
\begin{center}
\includegraphics[scale=0.38]{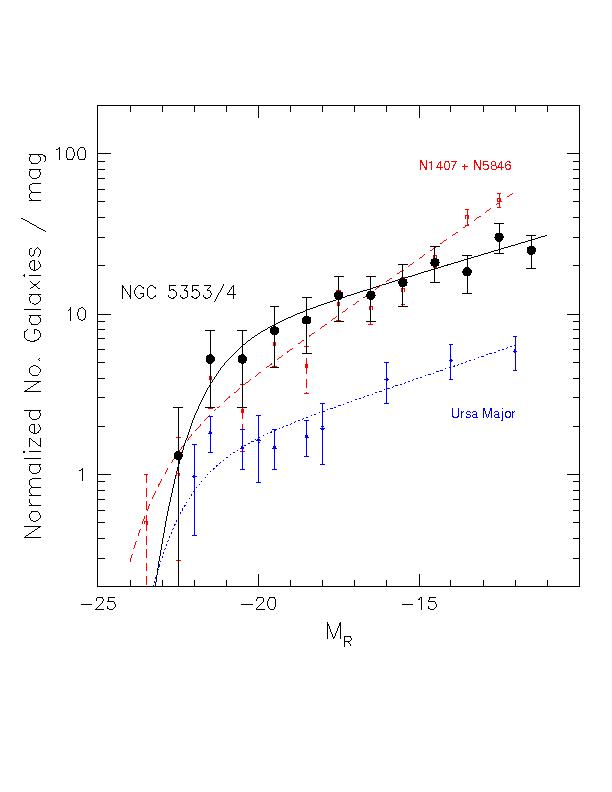}
\caption{Comparision of NGC~5353/4 luminosity function with equivalently normalized luminosity functions for the combined NGC~1407 and NGC~5846 groups (dashed red curve extracted from TTM06) and for the Ursa Major Cluster (dotted blue curve adapted from \citet{2001MNRAS.325..385T}). Error bars illustrate $\sqrt N / N$ variations within bins.}
\label{LF2}
\end{center}
\end{figure}

\begin{figure}[htbp]
\begin{center}
\includegraphics[scale=0.74]{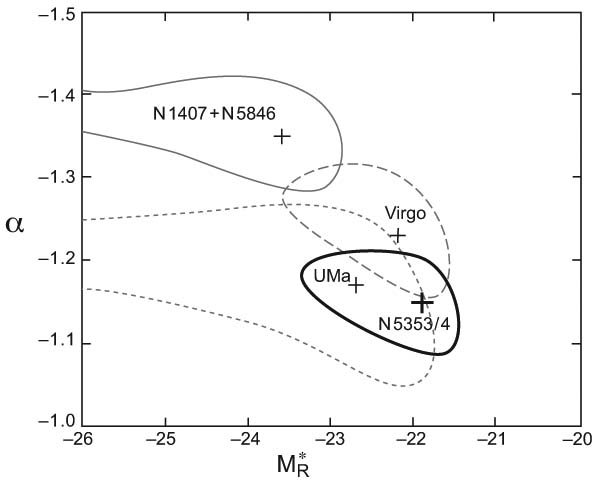}
\caption{Best values and 95\% probability limits for the Schechter function faint end slope $\alpha$ and the bright end exponential cutoff $M_R^{\ast}$. Within the 95\% probability limit, the $\chi^2$ fit is within $2\sigma$ of the best value, where $\sigma = (2/N)^{1/2}$ with $N=126$ galaxies in the sample with $M_R<-11$.  Fits for the NGC 5353/4 Group are shown by the heavy lines.  Similar information is plotted for other groups studied in the course of this program.  The exponential cutoff parameter is poorly constrained on the high luminosity side in groups with only a small number of luminous members.}
\label{alfa_mstar}
\end{center}
\end{figure}

The Schechter function fit \citep{1976ApJ...203..297S} shown by the solid curve in Fig.~\ref{LF} is characterized by the bright end exponential cutoff at $M_R^{\ast}=-21.9_{-0.6}^{+0.3}$ and the faint end power law slope $\alpha = -1.15 \pm 0.03$ ($1\sigma$ errors).   In Figure~\ref{LF2} the luminosity function found with this study is compared with similarly normalized luminosity functions found in our earlier work with similar observational material and analysis.   Figure~\ref{alfa_mstar} is an update of a figure in TTM06 and inter compares Schechter $M_R^{\ast}$ and $\alpha$ parameters for the separate groups that have been studied.
The Ursa Major Cluster \citep{2001MNRAS.325..385T} has a similar faint end slope and compatible bright end cutoff but has a distinctly lower normalization. 
The luminosity function for the combined NGC~1407 and NGC~5846 groups (TTM06) has distinctly fewer bright galaxies (aside from the central dominant systems NGC~1407 and the dumbbell pair NGC~5813 and NGC~5846) and more faint galaxies.

Could the luminosity function be relatively flat because group members are missed at low luminosities?  We note that the luminosity function seen in Fig.~\ref{LF} is constructed assuming {\it all} rating 3 systems are group members.  Within the framework of our sample this assumption maximally boosts the faint end of the luminosity function.

Still, could our sample be incomplete because high surface brightness objects were excluded by our OCP/ICP filter?  The possibility is highlighted by the spectroscopic discovery of 6 group members of this nature with $-17.5 < M_R < -14.5$.   Spectroscopic observations in other dense groups have revealed the existence of Ultra Compact Dwarfs \citep{2001ApJ...560..201P} which certainly are missed in the current survey.  However, if it were to be entertained that the faint end slope is as steep as given by $\alpha=-1.35$ found averaging the NGC~1407 and NGC~5846 groups then rather than the 94 galaxies found in the range $-17 < M_R < -11$ there should be 230; we would be missing $\sim 135$.  It is unlikely that a hypothetical compact population would so dominate the observed low surface brightness population.    Indeed, the search for Ultra Compact Dwarfs in groups by  \citet{2007MNRAS.378.1036E} indicates that these objects are not major constituents.  In the case of the NGC~1407 Group, those authors found 2 plausible (but unlikely) candidates in the magnitude range $17.5<b_j<20.5$, a range with 77 normal dwarf candidates in the catalog by TTM06.

An even stronger statement can be made regarding the {\it differential} between the NGC~5353/4 Group and the NGC~1407 and NGC~5846 groups.  Recall that the groups are at similar distances and have been observed in similar ways.  It is difficult to imagine that there would be a numerically dominant class of compact objects that escape detection in the NGC~5353/4 Group that does not exist in the NGC~1407 and NGC~5846 groups. The conclusion is inescapable that the luminosity function in the NGC~5353/4 Group is flatter than in the dynamically evolved knots of early-type systems around NGC~1407 and NGC~5846.  Parenthetically, our new result is similar to the slope $\alpha = 1.1$  \citet{2007A&A...463..503M} find in the case of the Fornax Cluster.  This is puzzling because Fornax is more similar in mass, $M/L$, and morphology properties to the NGC~1407 and NGC~5846 entities, yet seems to share the luminosity function properties of the NGC~5353/4 Group.

Looking back at Fig.~\ref{types}, it is apparent that the dwarfs that do exist in the NGC~5353/4 Group
are associated, for the most part, with the central region.  An individual dwarf cannot reasonably be related to a specific giant galaxy but, rather, simply to the central concentration.  There is very little concentration of the dwarfs in the vicinity of the large galaxies at the periphery of the group.  In particular, there is no noteworthy enhancement of dwarfs around NGC~5371, the giant spiral that is the brightest galaxy in the  group.

\section{Discussion and Summary}

Table~3 provides a summary of the properties of the NGC~5353/4 Group and, for comparison, the properties of the groups studied earlier in this series.

There have been two surprises for us in this study.  When we began, we were looking for a nearby region that complemented the Ursa Major Cluster by having numerous spiral galaxies, though at higher density in a smaller volume.  We were aware that the region that we chose contains several luminous S0 systems and intermediate luminosity E systems but the fact that 4 of these (and one spiral) are close together did not register as important to us.  The first surprise, then, was to see the pronounced concentration of the dwarf population around the four early-type galaxies.  The distribution of these mostly dE dwarf galaxies is clearly locating the dynamical center of the group to be around the pair of S0 galaxies NGC~5353 and NGC~5354.

\begin{figure}[htbp]
\begin{center}
\includegraphics[scale=0.38]{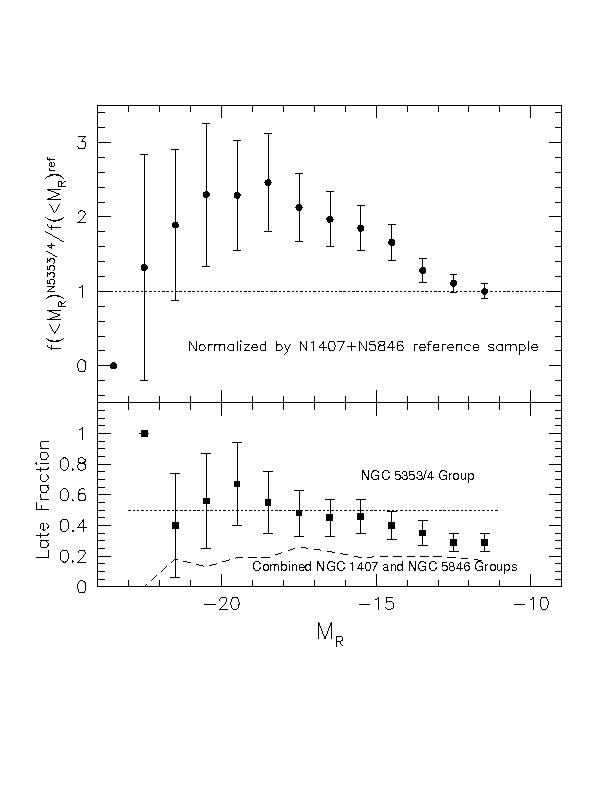}
\caption{Upper panel: Cumulative fraction of the total population of group candidates brighter than $M_R = -11$ normalized by the equivalent parameter for the combined NGC~1407 and NGC~5846 group populations.  The cumulative fractions reach unity at $M_R = -11$  so, by construction, the normalized function also reaches unity at $M_R = -11$.   Lower panel: Cumulative fraction of candidates with types later than Sa.  Points with error bars: the NGC~5353/4 Group sample.  Dashed curve: the combined NGC~1407 and NGC~5846 groups sample.}
\label{mag_cum}
\end{center}
\end{figure}

Overall, the group luminosity function is quite flat at the faint end.  It is as flat as seen in the spiral rich Ursa Major Cluster and significantly flatter than seen in the E/S0 dominated NGC~5846 and NGC~1407 groups (see Figs.~\ref{LF2} and \ref{alfa_mstar}).  Part of the story is at the luminous end as can be seen with Figure~\ref{mag_cum}.   NGC~1407 is a very luminous galaxy, almost 5 times brighter than the next brightest in the group, and the dominant pair NGC~5846 and NGC~5813 in the NGC~5846 Group are both more luminous than the brightest galaxy in the NGC~5353/4 Group.  Thereafter, though, there are a lot more giant and intermediate luminosity systems in the region of the current study than around NGC~5846 or NGC~1407.

We define a cumulative fraction function 
\begin{equation}
f(<M_R) = N(<M_R) / N(<-11)
\end{equation} 
and normalize the function derived for the NGC~5353/4 Group by a reference function constructed from  the combined NGC 1407 and NGC 5846 groups:  $f(<M_R)^{N5353/4} / f(<M_R)^{N1407+N5846}$ where $N(<M_R)$ is the number of
group members brighter than $M_R$ and $N(<-11)$ is the number brighter than $M_R = -11$.  The normalized cumulative fraction is seen in Fig.~\ref{mag_cum} to peak in the $M_R=-18$ bin.  Relative to the two groups studied in detail earlier, the NGC~5353/4 Group has a substantially larger fraction of its members in the magnitude range $-22<M_R<-18$ and/or fewer dwarfs.  There is also an obvious difference in the morphologies of the higher luminosity systems in these separate groups.  The NGC~5353/4 Group maintains greater than or near 50\% late types in all magnitude bins brighter than $M_R=-15$, peaking at 2/3 in the cumulative count in the $-19$ bin, before falling to 30\% in the overall sample at $M_R=-11$.  By contrast, the combined sample of the NGC~1407 and NGC~5846 groups holds at a late fraction of 20\% at all
magnitudes. 

Even if one just considers the early types in the NGC~5353/4 Group there are fewer dwarfs per giant than in the NGC~5846 and NGC~1407 groups.  If one considers just the late types in the NGC~5353/4 Group the deficiency of dwarfs per giant is extreme.

\begin{figure}[htbp]
\begin{center}
\includegraphics[scale=0.38]{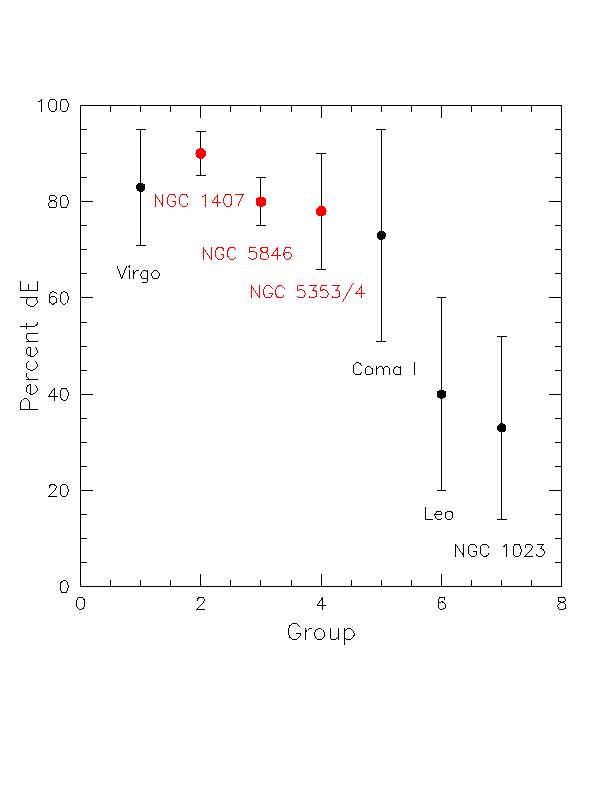}
\caption{Percentage of dwarfs with $-17 < M_R < -11$ that are early type.  The dE/I transition objects
are split equally between early and late for the purposes of this plot.  Data for other groups are from 
TTM06 and other papers in this series.}
\label{pcE}
\end{center}
\end{figure}

The overall morphological typing of the dwarfs in the NGC~5353/4 Group is not so different from that in the E/S0 dominated NGC~5846 and NGC~1407 groups.  Figure~\ref{pcE} has our new data added to the material of TTM06.  The dwarf population in the current study is predominantly of early (dE = dwarf elliptical) type; not dI (dwarf irregular) or BCD (blue compact dwarf) or VLSB (very low surface brightness). 

We find that the NGC~5353/4 Group is highly segregated.  The more luminous galaxies are predominantly spiral galaxies which show a preference for the suburbs.  There are very few irregular dwarfs.  Most dwarfs are dE and they are concentrated toward the downtown core.

The second surprise for us was to see from the SDSS spectra that many of those dE and other early types are busy forming stars.  These are galaxies that appear old and relaxed.  Even so, gas has accreted recently at their centers and hot stars have been born.  It is seen from Figs. \ref{types} and \ref{cumulative} that the particular `transition' or star forming early type galaxies that have been found tend to be at intermediate radii in the group.  They are {\it not} so centrally concentrated as the overall dE class.  It is true that the ones with spectra are among the more luminous dwarfs and to a degree may reflect the trend for more luminous objects to be more dispersed.  At any rate, while the overall dE class of objects in the NGC~5353/4 Group may be relatively relaxed around the central galaxies, we conclude that the fraction manifesting star formation are {\it not} relaxed within the group potential.

The contrast with the two groups studied earlier in this series is striking.  The NGC~1407 and NGC~5846 groups contain overwhelmingly early type galaxies.  Only modest signs of star formation have been found in the members.  It was argued by TTM06 that the NGC~1407 Group, with a single dominant very luminous E galaxy,  is dynamically very old, related to the `fossil group' class \citep{1994Natur.369..462P}.  The NGC~5846 Group is taken to be somewhat less dynamically evolved; it is a dumbbell system with two dominant and well separated E galaxies (MTT05).  We view the NGC~5353/4 Group as an interesting intermediate case in the sequence of dynamical evolution.  There is a distinct core of early types with an impending major merger (between NGC~5353 and NGC~5354).  There is a healthy reservoir of large late types available for further merging and growth of the central system.

This series emphasizes the importance of dwarf galaxies as dynamical markers.  Dwarf galaxies of type dE are strongly clustered around major E/S0 hosts.  The motions of galaxies show that a lot of matter is located in these regions.  Given the amount of light, most of the matter is `dark'.  Yet in the present case there is something curious.  The ensemble group has $M/L_R = 105 ~\ML$.  If only the light of the early type galaxies is considered then $M/L_R^{E/S0} \sim 225$.  This value is significantly lower than the values for the NGC~5846 Group ($M/L_R = 320 \ML$) and NGC~1407 Group ($M/L_R = 340 \ML$).  If the late type galaxies are ultimately absorbed in their entirety and the entire stellar population is old, the old population luminosity will be augmented and the group $M/L_R$ value would be intermediate between the present $105~\ML$ and $225~\ML$; i.e., perhaps a factor 2 less than found in the two very evolved groups around NGC~1407 and NGC~5846.  Plausibly, in the collisions yet to be experienced in the NGC~5353/4 Group a lot of stars will be thrown out of galaxies (the light inventory in the $M/L_R$ calculation only counts the light in galaxies).  Alternatively, perhaps a larger fraction of baryons in the evolved NGC~1407 and NGC~5846 groups are in intragroup plasmas not hot enough to register at X-ray bands.  At any rate, swarms of dE appear to occur around dominant E/S0 where $M/L$ values are relatively high and are not seen in large numbers in groups dominated by spirals where $M/L$ values are relatively low ( \citet{2005ApJ...618..214T} discusses $M/L$ variations with environment).  As a local example, Cen~A is the host of many more dE than the other nearby giants \citep{2007AJ....133..504K}.
{\it Congregations of dwarf galaxies of type dE appear to be markers for regions of concentrated dark matter.}

We also have learned that dE galaxies can harbor surprises.  Star formation can occur at the centers of galaxies that are textureless and apparently relaxed on large scales.  These properties are reminiscent  of conditions in dwarfs around the Milky Way.  Most have experienced multiple periods of star formation, continuing until recently \citep{1998ARA&A..36..435M}.  This information about the dwarfs in the NGC~5353/4 Group, and the contrast with those around NGC~1407 and NGC~5846, is consistent with the proposition that the NGC~5353/4 Group is at an intermediate evolutionary stage.  \citet{1980ApJ...237..692L} proposed that the time scale for star formation to exhaust gas reservoirs in galaxies is generally less than a Hubble time, requiring that there be a replenishment source for systems displaying activity today.  Those authors proposed that tidal interactions could disrupt the acquisition of gas onto galaxies in clusters and account for the correlation between galaxy type and environment.     \citet{1984ApJ...281...56S} went further by pointing out that a cluster potential imposes a Roche limit around a galaxy.  Once a galaxy enters a cluster, shells of gas that have not yet accreted are deposited into the larger potential well of the cluster and virialized at a high temperature.  This process has come to be called `starvation'.  This mechanism may have relevance in the case of the NGC~5353/4 Group because it is necessary to explain why there are patterns of star formation on scales of hundreds of kiloparsecs within the group.  Perhaps these patterns are a relic of the group infall history.  Gas starvation will occur over several times the dynamical group crossing time of $r_{2t}/(3^{1/2}\sigma_V) \sim 1.4$~Gyr.  Perhaps more recent infall into the NGC~5353/4 Group has deposited galaxies preferentially in the south--west quadrant where there is a high incidence of ongoing star formation, somewhat earlier infall distributed galaxies into the east--west band exhibiting older star formation characteristics, and the infall into the NGC~5846 Group was earlier still. Accordingly, star formation in these modest galaxies would be a sputtering process not strongly coupled to external influences but more governed by a diminishing availability of gas.

It has been suggested that dI transform to dE by ram pressure stripping \citep{1972ApJ...176....1G} or harassment in groups \citep{1996Natur.379..613M}.  There is no evidence of a sufficient intragroup plasma for the former mechanism to be effective in the NGC~5353/4 Group, and the dispersed distribution of dwarfs with respect to big galaxies in Fig.~\ref{types} gives no succor to the proposition that harassment has initiated the many star formation episodes.  Perhaps greater attention should be given to the `starvation' mechanism.  We offer another piece of evidence on the matter.  \citet{2003AJ....125.1926G} have noted that at a given absolute magnitude dE galaxies have significantly higher metallicities than dI.   Sharina et al. (2007) show convincing evidence for this effect in nearby dwarfs.  If dI transform to dE and the metallicity--luminosity dichotomy is explained solely by luminosity dimming then 90-99\% of $B$ light is routinely lost in the transition.  The implications are less extreme if a significant part of the transition is in metallicity.  The general trend to lower metallicities in smaller galaxies can be understood to be a consequence of energetic events causing greater mass loss in smaller halos  \citep{1986ApJ...303...39D}.  However, consider the distinctive environments within and outside of groups.  A dwarf galaxy inside a group will blow off interstellar material during star formation episodes but, deprived of infalling gas, the mean metallicity of remaining gas will rise.  By contrast, a dwarf outside of a group will have continued to accrete pristine gas from its protogalaxy reservoir until relatively recently.  There will be competition between metal enhanced outflow and pristine inflow that keeps metal abundances reduced.

The undramatic starvation mechanism must work at some level and it is just a question of its importance relative to other mechanisms.  According to this idea, dwarf galaxies outside of groups have accreted fresh gas until relatively recently, like their giant counterparts.  Once such a galaxy falls into a group it has only its internal resources to draw upon and after a few group crossing times those resouces are depleted.  The situation in the NGC~5353/4 Group favors this starvation mechanism as an explanation for the putative dI to dE transformation.
{\it The spatial displacements of the systems with greater and lesser activity may be signatures of the pattern of recent arrivals to the group.}

Features in the luminosity function might be indicators of dynamical evolution.  A local minimum or concave inflection at intermediate luminosities is seen in some environments.   \citet{2004MNRAS.355..785M} identify such a feature in groups with low X-ray luminosity and argue that this is evidence for depletion of intermediate luminosity systems through merging in small groups with low velocity dispersions.  MTT05 made a similar identification in the case of the NGC~5846 Group.  Ironically, our NGC~5353/4 Group, where we do not see any indication of an inflection, is in the Miles et al. low X-ray luminosity sample that collectively shows the effect and the NGC~5846 Group, where we see the luminosity dip, is in the Miles et al. high X-ray luminosity sample that collectively shows no such effect.  Little should be made of these exceptions.  The larger point is that luminosity function shapes do vary and dynamical evolution provides a plausible explanation.  In the case under consideration, it is probable that a substantial number of gas-rich intermediate luminosity systems has rained in on a previously formed potential well that we find delineated by the clustering of dE around an E/S0 knot.

Overall, the formation timescale for the NGC 5353/4 Group is thought to be significantly longer than for the two previously studied groups.  While there is more star formation activity in the NGC~5353/4 dwarfs, there are fewer of them compared with the number of giants. The abundance of low luminosity dwarfs might be regulated by the joint timescales of the epoch of group formation and reionization \citep{2002ApJ...569..573T}.  It is expected that low mass dark halos collapsed earlier in denser environments with the consequence that in these places more protodwarfs were in place with gas and stars before the epoch of reionization.
{\it The flatter faint end slope of the NGC~5353/4 Group luminosity function might be a signature of an intermediate evolutionary age.}

\acknowledgments
An exceptionally diligent anonymous referee has led to the substantial improvement of the presentation of our paper.  This research involved observations with the MegaCam imager on the Canada--France--Hawaii Telescope.
Extensive use was made of the Sloan Digital Sky Survey archive and the NASA-IPAC Extragalactic Database,  Support from the US National Science Foundation was provided by grant AST 03-07706.

Images of the galaxies identified in Table 1 can be obtained from the CFHT cutout service: 
http://www.cadc.hia.nrc.gc.ca/cadcbin/cfhtCutout

\bibliography{paper}

\begin{deluxetable}{rllcccccccc}
\tablecaption{Members and Candidate Members of the NGC 5353/4 Group}
\label{catalog}
\tablewidth{0in}
\tablehead{\colhead{N5353--} & \colhead{Name} & \colhead{Type} & \colhead{Rate} & \colhead{$V_h$ (\kms)} & \colhead{$\alpha$ (J2000)} & \colhead{$\delta$ (J2000)} & \colhead{$r$ (deg)} & \colhead{$R_{350}$} & \colhead{$R$} & \colhead{$M_R$}}
\startdata
 \multicolumn{11}{l}{Candidates within radius of $1.23^{\circ} = 625$~kpc -- ordered by luminosity} \\
 \hline
    1  &  NGC 5371  &  Sbc     &  0  &  2558  &  13 55 40.0  &   40 27 42  & 0.4560  &  14.27  &  10.04  &  -22.30 \\
   2  &  NGC 5353  &  S0      &  0  &  2325  &  13 53 26.7  &   40 16 59  & 0.0099  &  14.22  &  10.57  &  -21.77 \\
   3  &  NGC 5354  &  S0      &  0  &  2570  &  13 53 26.7  &   40 18 10  & 0.0098  &  14.14  &  10.64  &  -21.70 \\
   4  &  NGC 5350  &  Sb      &  0  &  2321  &  13 53 21.6  &   40 21 50  & 0.0727  &  14.76  &  10.99  &  -21.35 \\
   5  &  NGC 5311  &  S0/a    &  0  &  2398  &  13 48 56.1  &   39 59 06  & 0.9135  &  13.58  &  11.09  &  -21.26 \\
   6  &  NGC 5326  &  Sa      &  0  &  2520  &  13 50 50.7  &   39 34 31  & 0.8723  &  13.63  &  11.38  &  -20.97 \\
   7  &  NGC 5313  &  Sb      &  0  &  2548  &  13 49 44.3  &   39 59 05  & 0.7712  &  14.54  &  11.43  &  -20.90 \\
   8  &  NGC 5320  &  Sc      &  0  &  2619  &  13 50 20.4  &   41 21 58  & 1.2256  &  16.03  &  11.98  &  -20.35 \\
   9  &  NGC 5362  &  Sb      &  0  &  2175  &  13 54 53.3  &   41 18 49  & 1.0571  &  15.83  &  12.17  &  -20.16 \\
  10  &  NGC 5337  &  Sab     &  0  &  2165  &  13 52 23.0  &   39 41 14  & 0.6387  &  15.26  &  12.42  &  -19.93 \\
  11  &  UGC 8736  &  Sc      &  0  &  2384  &  13 49 04.4  &   39 29 57  & 1.1512  &  16.68  &  13.01  &  -19.34 \\
  12  &  NGC 5355  &  E       &  0  &  2355  &  13 53 45.6  &   40 20 20  & 0.0756  &  15.14  &  13.08  &  -19.26 \\
  13  &  UGC 8726  &  Sd      &  0  &  2334  &  13 47 48.8  &   40 29 15  & 1.0914  &  18.43  &  13.08  &  -19.25 \\
  14  &  IC 4336   &  Sb      &  0  &  2503  &  13 50 43.6  &   39 42 22  & 0.7831  &  17.17  &  13.12  &  -19.24 \\
  15  &  NGC 5346  &  Scd     &  0  &  2519  &  13 53 01.9  &   39 34 50  & 0.7168  &  17.10  &  13.18  &  -19.17 \\
    16  &  NGC 5358  &  S0      &  0  &  2401  &  13 54 00.4  &   40 16 39  & 0.1082  &  15.20  &  13.35  &  -18.99 \\
  17  &  CGCG 218$-$060  &  E       &  0  &  2313  &  13 50 41.4  &   40 16 45  &   0.5256 & 15.60  &  13.70  &  -18.63 \\
  18  &  UGC 8841  &  Sc      &  0  &  2422  &  13 55 08.0  &   40 10 02  & 0.3456  & 17.64  &  13.79  &  -18.56 \\
  19  &  Mrk 462   &  S0/a    &  0  &  2355  &  13 51 25.4  &   40 12 48  & 0.3937  &  15.83  &  13.83  &  -18.51 \\
  20  &            &  Sdm     &  0  &  2310  &  13 47 50.1  &   40 29 00 & 1.0866 &  18.40  &  13.86  &  -18.47 \\
  21  &  KUG 1351+402  &  Sa      &  0  &  2622  &  13 54 03.8  &   39 59 07  & 0.3295 &  17.05  &  14.02  &  -18.33 \\
  22  &  UGC 8840  &  dE      &  0  &  2194  &  13 55 07.6  &   39 56 10  &  0.4798  & 16.65  &  14.07  &  -18.28 \\
  23  &  UGC 8721  &  Sdm     &  0  &  2617  &  13 47 33.7  &   40 21 04  &  1.1234 & 18.25  &  14.42  &  -17.91 \\
  24  &  CGCG 219$-$021  &  S0      &  0  &  2613  &  13 53 59.4  &   39 42 56  &  0.5867 & 16.86  &  14.44  &  -17.91 \\
  25  &  UGC 8807  &  Sm      &  0  &  1968  &  13 53 16.2  &   40 47 46  & 0.5023 & 17.61  &  14.68  &  -17.65 \\
  26  &            &  dE/I    &  0  &  2228  &  13 53 40.3  &   39 43 27  &  0.5705 &  16.61  &  14.76  &  -17.59 \\
  27  &  NPM1G+40.0334  &  S0/a    &  0  &  2543  &  13 49 38.0  &   40 03 48  & 0.7623 &  17.11  &  14.92  &  -17.41 \\
  28  &  KUG 1347+399  &  S0/a    &  5  &  2758  &  13 49 45.5  &   39 42 51  & 0.9107 & 17.33  &  14.93  &  -17.40 \\
  29  &            &  S0/a    &  3  &  2698  &  13 52 24.8  &   39 33 26  & 0.7616 & 16.75  &  14.96  &  -17.39 \\
  30  &  KUG 1350+399  &  Sm      &  0  &  2633  &  13 52 58.4  &   39 43 07  &  0.5814  & 17.41  &  14.96  &  -17.39 \\
  31  &            &  E       &  3  &  2248  &  13 58 44.3  &   40 54 50  & 1.1850 &  17.39  &  14.98  &  -17.36 \\
  32  &            &  E/dE    &  3  &  2728  &  13 54 07.6  &   40 49 01  & 0.5398 &  17.14  &  15.11  &  -17.22 \\
  33  &            &  S0/a    &  3  &  2188  &  13 56 50.9  &   39 56 16  & 0.7398 &  18.40  &  15.39  &  -16.96 \\
  34  &            &  dE/I    &  3  &  2368  &  13 54 55.1  &   39 58 39  & 0.4224 &  18.86  &  15.50  &  -16.85 \\
  35  &            &  E/dE   &  3  &        &  13 52 41.3  &   40 11 41  & 0.1746 &  17.76  &  15.76  &  -16.58 \\
  36  &            &  E       &  3  &  2638  &  13 51 05.2  &   39 22 09  & 1.0275 &  18.07  &  15.86  &  -16.49 \\
  37  &  LEDA 099754  &  E       &  0  &  2597  &  13 53 18.8  &   40 19 01   &  0.0347  &  17.99  &  15.87  &  -16.47 \\
  38  &            &  dE/I    &  5  &  2578  &  13 51 03.7  &   40 25 51  &  0.4749 & 16.97  &  15.89  &  -16.44 \\
  39  &            &  Sm      &  2  &  2218  &  13 50 42.7  &   40 02 08  &  0.5814 & 18.69  &  15.91  &  -16.43 \\
  40  &            &  dI      &  2  &  2458  &  13 52 12.4  &   40 05 33  & 0.3098 &  19.12  &  15.99  &  -16.35 \\
  41  &            &  dE/I    &  5  &  2848  &  13 53 40.5  &   41 15 08  & 0.9602 &  17.72  &  16.18  &  -16.15 \\
  42  &            &  E       &  3  &  2248  &  13 55 27.5  &   41 08 20  & 0.9289 &  18.53  &  16.18  &  -16.15 \\
  43  &            &  dE/I    &  3  &  2728  &  13 54 14.1  &   40 28 38  & 0.2379 &  18.57  &  16.34  &  -15.99 \\
  44  &            &  E/dE    &  3  &        &  13 52 10.6  &   40 18 15  & 0.2422 &  18.49  &  16.50  &  -15.84 \\
  45  &            &  dE/I    &  2  &  2698  &  13 55 04.8  &   40 34 18  & 0.4181 &  19.06  &  16.52  &  -15.82 \\
  46  &            &  dE/I    &  5  &  2578  &  13 48 58.8  &   40 22 33  & 0.8555 & 18.37  &  16.57  &  -15.76 \\
  47  &            &  Sm      &  3  &  2158  &  13 54 53.5  &   41 18 30  & 1.0521 & 17.73  &  16.65  &  -15.68 \\
  48  &            &  Sm     &  5  &  1979  &  13 55 27.7  &  39 58 47  & 0.4960 & 18.59  &  16.75  &  -15.59 \\
  49  &            &  dE      &  1  &        &  13 56 09.8  &   40 18 05  & 0.5184 & 21.11  &  16.78  &  -15.57 \\
  50  &            &  dE/I    &  3  &  2608  &  13 51 53.9  &   41 18 00  & 1.0493 & 18.52  &  16.82  &  -15.50 \\
  51  &  LEDA 166181  &  Sm      &  0  &  2593  &  13 48 14.8  &   40 33 09  &  1.0247   &  20.69  &  16.91  &  -15.42 \\
  52  &            &  dE      &  3  &  2248  &  13 50 24.3  &   39 51 29  & 0.7248 & 19.18  &  17.14  &  -15.20 \\
  53  &            &  dE      &  3  &  2026  &  13 51 39.3  &   40 44 14  & 0.5602 & 19.23  &  17.26  &  -15.07 \\
  54  &            &  dI      &  3  &        &  13 53 02.2  &   40 05 41  & 0.2130 & 20.21  &  17.33  &  -15.01 \\
  55  &            &  E/dE    &  3  &        &  13 53 12.7  &   40 14 37  & 0.0665 & 18.97  &  17.35  &  -14.99 \\
  56  &            &  dE      &  3  &        &  13 51 19.9  &   40 11 32 & 0.4154 &  19.91  &  17.38  &  -14.96 \\
  57  &           &  dI      &  3  &  2398  &  13 51 26.2  &   41 11 05  & 0.9705 &  19.20  &  17.38  &  -14.94 \\
  58  &            &  dE/I    &  5  &  2460  &  13 50 23.2  &   41 06 21  & 1.0004 &  18.14  &  17.51  &  -14.82 \\
  59  &            &  dE      &  0  &  2428  &  13 49 11.0  &   40 19 37  & 0.8134 &  19.93  &  17.64  &  -14.69 \\
  60  &            &  dE,N    &  3  &        &  13 52 58.1  &   40 16 06  & 0.0942 &  19.67  &  17.68  &  -14.66 \\
  61  &            &  VLSB    &  2  &        &  13 53 25.0  &   40 19 46  & 0.0368 &  21.68  &  17.75  &  -14.59 \\
  62  &            &  dE      &  3  &        &  13 55 48.7  &   40 34 14  & 0.5298 &  19.67  &  17.77  &  -14.57 \\
  63  &            &  dE,N/I  &  2  &        &  13 49 20.1  &   39 57 09  & 0.8545 &  20.95  &  17.82  &  -14.52 \\
  64  &            &  dE      &  2  &        &  13 57 36.0  &   39 46 45  & 0.9443 &  20.95  &  17.91  &  -14.44 \\
  65  &            &  dE      &  3  &        &  13 53 00.2  &   40 16 54  & 0.0850 &  19.68  &  17.92  &  -14.42 \\
  66  &            &  dE      &  3  &        &  13 54 30.5  &   41 26 52  & 1.1724 &  20.02  &  17.93  &  -14.39 \\
  67  &            &  dE,N    &  1  &        &  13 52 00.6  &   40 22 35  & 0.2861 &  20.91  &  17.99  &  -14.35 \\
  68  &            &  E/dE    &  3  &        &  13 54 52.7  &   39 57 56  & 0.4265 &  20.16  &  18.01  &  -14.34 \\
  69  &            &  dE,N    &  1  &        &  13 52 28.7  &   40 24 30  & 0.2175 &  21.41  &  18.25  &  -14.09 \\
  70  &            &  dE      &  3  &        &  13 53 41.9  &   40 04 48  & 0.2184 &  20.20  &  18.33  &  -14.01 \\
  71  &            &  dE      &  3  &        &  13 54 08.2  &   40 20 05  & 0.1383 &  20.17  &  18.48  &  -13.85 \\
  72  &            &  dE      &  3  &        &  13 57 10.5  &   40 11 42  & 0.7180 &  20.16  &  18.54  &  -13.80 \\
  73  &            &  dE      &  3  &        &  13 56 34.6  &   41 01 04  & 0.9391 &  21.45  &  18.60  &  -13.72 \\
  74  &            &  dI      &  3  &        &  13 52 53.1  &   39 41 04  & 0.6179 &  20.89  &  18.77  &  -13.58 \\
  75  &            &  dE      &  3  &        &  13 57 37.2  &   39 34 32  & 1.0717 &  20.37  &  18.89  &  -13.47 \\
  76  &            &  dE      &  3  &        &  13 54 53.8  &   40 26 48  & 0.3166 &  20.42  &  18.91  &  -13.42 \\
  77  &            &  E/dE    &  3  &        &  13 53 58.6  &   40 12 48  & 0.1289 &  20.28  &  19.02  &  -13.33 \\
  78  &            &  dE      &  3  &        &  13 49 53.2  &   40 00 13  & 0.7377 &  20.68  &  19.03  &  -13.30 \\
  79  &            &  dE      &  3  &        &  13 50 14.9  &   40 14 12  & 0.6122 &  20.51  &  19.05  &  -13.28 \\
  80  &            &  dE      &  3  &        &  13 52 42.2  &   40 02 05  & 0.2945 &  20.67  &  19.07  &  -13.27 \\
  81  &            &  dE,N    &  3  &        &  13 51 26.3  &   40 35 10  & 0.4820 &  21.08  &  19.11  &  -13.22 \\
  82  &            &  dE      &  3  &        &  13 54 55.6  &   39 49 13  & 0.5507 &  20.74  &  19.15  &  -13.20 \\
  83  &            &  dE/I    &  3  &        &  13 50 12.4  &   39 57 18  & 0.7040 &  21.53  &  19.19  &  -13.14 \\
  84  &            &  dE      &  3  &        &  13 51 40.4  &   40 08 56  & 0.3673 &  20.97  &  19.24  &  -13.10 \\
  85  &            &  dE,N    &  3  &        &  13 52 41.8  &   39 49 58  & 0.4819 &  20.73  &  19.41  &  -12.93 \\
  86  &            &  dE      &  3  &        &  13 52 53.5  &   40 25 22  & 0.1673 &  20.63  &  19.43  &  -12.91 \\
  87  &            &  dE      &  2  &        &  13 55 43.2  &   40 33 22  & 0.5073 &  21.03  &  19.48  &  -12.86 \\
  88  &            &  dE,N    &  2  &        &  13 53 02.3  &   40 17 46  & 0.0776 &  20.75  &  19.57  &  -12.77 \\
  89  &            &  dE      &  3  &        &  13 55 17.8  &   40 08 40  & 0.3830 &  20.99  &  19.59  &  -12.76 \\
  90  &            &  dE      &  3  &        &  13 55 00.7  &   40 03 18  & 0.3819 &  21.14  &  19.61  &  -12.73 \\
  91  &            &  dE      &  3  &        &  13 57 16.7  &   40 12 59  & 0.7349 &  21.26  &  19.62  &  -12.72 \\
  92  &            &  dE,N    &  3  &        &  13 52 19.9  &   40 19 35  & 0.2149 &  20.53  &  19.67  &  -12.67 \\
  93  &            &  dE,N    &  3  &        &  13 54 05.4  &   40 13 49  & 0.1380 &  20.91  &  19.68  &  -12.66 \\
  94  &            &  dE,N    &  2  &        &  13 51 20.4  &   39 37 44  & 0.7760 &  20.96  &  19.70  &  -12.64 \\
  95  &            &  dE      &  3  &        &  13 51 05.2  &   40 25 07  & 0.4469 &  21.25  &  19.90  &  -12.43 \\
  96  &            &  dE      &  2  &        &  13 55 26.3  &   40 16 24  & 0.3806 &  21.77  &  19.91  &  -12.43 \\
  97  &            &  dE,N    &  3  &        &  13 54 08.4  &   40 03 51  & 0.2644 &  21.47  &  19.92  &  -12.42 \\
  98  &            &  dE,N    &  3  &        &  13 51 08.8  &   39 41 46  & 0.7405 &  21.48  &  19.95  &  -12.40 \\
  99  &            &  dE,N    &  3  &        &  13 52 58.2  &   40 07 02  & 0.1978 &  21.03  &  20.05  &  -12.29 \\
 100  &            &  dI      &  3  &        &  13 54 10.4  &   40 12 03  & 0.1666 &  21.79  &  20.06  &  -12.28 \\
 101  &            &  dE,N    &  3  &        &  13 58 08.0  &   39 55 40  & 0.9657 &  21.03  &  20.08  &  -12.27 \\
 102  &            &  dE      &  2  &        &  13 53 15.9  &   40 27 57  & 0.1762 &  21.91  &  20.09  &  -12.25 \\
 103  &            &  dE      &  3  &        &  13 51 22.3  &   40 10 42  & 0.4117 &  21.94  &  20.10  &  -12.24 \\
 104  &            &  dI      &  3  &        &  13 53 51.3  &   40 16 15  & 0.0812 &  22.04  &  20.12  &  -12.22 \\
 105  &            &  dE      &  3  &        &  13 53 06.6  &   40 30 02  & 0.2172 &  21.95  &  20.27  &  -12.07 \\
 106  &            &  dE,N    &  2  &        &  13 54 45.3  &   40 14 12  & 0.2560 &  21.67  &  20.28  &  -12.05 \\
 107  &            &  dE      &  3  &        &  13 54 02.9  &   37 17 57  & 0.5955 &  21.70  &  20.34  &  -12.00 \\
 108  &            &  dE      &  3  &        &  13 52 06.5  &   40 19 43  & 0.2574 &  21.70  &  20.37  &  -11.97 \\
 109  &            &  dE/I    &  3  &        &  13 52 49.8  &   40 09 19  & 0.1809 &  21.65  &  20.40  &  -11.94 \\
 110  &            &  dE,N    &  2  &        &  13 58 17.0  &   40 11 54  & 0.9274 &  22.44  &  20.43  &  -11.91 \\ 
 111  &            &  dE      &  3  &        &  13 56 58.9  &   41 04 57  & 1.0383 &  22.12  &  20.54  &  -11.79 \\
 112  &            &  dE      &  2  &        &  13 51 17.8  &   39 29 30  & 0.9000 &  22.17  &  20.76  &  -11.59 \\
 113  &            &  dE      &  3  &        &  13 54 32.1  &   40 11 06  & 0.2342 &  21.77  &  20.77  &  -11.57 \\
 114  &            &  dI      &  3  &        &  13 56 10.3  &   39 15 52  & 1.1525 &  21.79  &  20.84  &  -11.53 \\
 115  &            &  dI      &  3  &        &  13 49 14.7  &   40 11 33  & 0.8072 &  22.21  &  20.81  &  -11.52 \\
 116  &            &  dE      &  2  &        &  13 51 17.0  &   40 31 02  & 0.4693 &  22.26  &  20.81  &  -11.52 \\
 117  &            &  dI      &  3  &        &  13 51 58.1  &   40 08 35  & 0.3190 &  21.85  &  20.85  &  -11.48 \\
 118  &            &  dE      &  3  &        &  13 55 37.5  &   40 14 59  & 0.4179 &  22.07  &  20.93  &  -11.41 \\
 119  &            &  dE      &  3  &        &  13 53 45.6  &   40 16 06  & 0.0649 &  23.44  &  20.95  &  -11.39 \\
 120  &            &  dE      &  3  &        &  13 54 06.3  &   40 08 39  & 0.1949 &  21.85  &  21.03  &  -11.31 \\
 121  &            &  dE      &  2  &        &  13 52 38.8  &   39 11 31  & 1.1115 &  22.29  &  21.09  &  -11.28 \\
 122  &            &  dE      &  3  &        &  13 48 07.6  &   40 19 25  & 1.0146 &  21.68  &  21.16  &  -11.17 \\
 123  &            &  dE/I    &  3  &        &  13 52 02.9  &   40 01 53  & 0.3734 &  22.21  &  21.20  &  -11.14 \\
 124  &            &  dE/I    &  2  &        &  13 56 19.3  &   40 24 40  & 0.5611 &  22.96  &  21.25  &  -11.09 \\
 125  &            &  dE      &  3  &        &  13 53 02.6  &   41 05 15  & 0.7982 &  22.29  &  21.28  &  -11.05 \\
 126  &            &  dE,N    &  3  &        &  13 53 59.5  &   39 36 04  & 0.6997 &  22.06  &  21.29  &  -11.05 \\
 127  &            &  dE      &  3  &        &  13 54 18.4  &   40 35 24  & 0.3394 & 22.51  &  21.39  &  -10.94 \\
 128  &            &  dE      &  2  &        &  13 52 23.1  &   40 20 23  & 0.2075  &  23.00  &  21.49  &  -10.85 \\
 129  &            &  dE      &  3  &        &  13 55 37.5  &   40 36 38  & 0.5231 &  21.92  &  21.56  &  -10.77 \\
 130  &            &  dE/I    &  3  &        &  13 50 20.3  &   40 13 20  & 0.5967 &  22.40  &  21.66  &  -10.67 \\
 131  &            &  dE/I    &  3  &        &  13 55 12.6  &   40 34 09  & 0.4353  &  22.83  &  21.69  &  -10.65 \\
 132  &            &  dE/I    &  3  &        &  13 49 51.6  &   40 00 16  & 0.7420 &  22.28  &  21.80  &  -10.53 \\
 133  &            &  dE/I    &  3  &        &  13 55 59.8  &   40 32 11  & 0.5440 & 22.74  &  21.91  &  -10.43 \\
 134  &            &  dE      &  3  &        &  13 55 04.4  &   40 36 19  & 0.4404 & 22.76  &  22.05  &   -10.28 \\
 135  &            &  dI      &  3  &        &  13 50 05.9  &   40 58 13  &  0.9306  & 22.81  &  22.08  &   -10.25 \\
 136  &            &  dE/I    &  3  &        &  13 55 15.2  &   40 07 04  &  0.3868  & 22.61  &  22.16  &   -10.19 \\
 137  &            &  dE      &  3  &        &  13 53 39.4  &   40 02 17  & 0.2581 &  22.89  &  22.69  &   ~-9.66 \\
\hline
\multicolumn{11}{l}{Candidates outside radius of $1.23^{\circ}$ -- ordered by distance from center of group} \\
\hline
 138  &            &  dI      &  3  &        &  13 58 26.3  &   39 26 55  &  1.2726   &  23.31  &  20.32  &  -12.04 \\
 139  &            &  dE      &  3  &        &  13 50 17.8  &   41 25 57  &  1.2880   &  21.44  &  20.20  &  -12.13 \\
 140  &            &  dI      &  0  &  2790  &  13 47 58.3  &   39 23 22  &  1.3805   &  18.71  &  16.46  &  -15.88 \\
 141  &            &  dE      &  3  &        &  13 52 02.8  &   38 56 15  &  1.3815   &  21.93  &  21.09  &  -11.28 \\
 142  &            &  dE      &  3  &        &  13 51 44.1  &   38 52 18  &  1.4583   &  22.56  &  20.80  &  -11.58 \\
 143  &            &  dE      &  2  &        &  13 55 13.6  &   41 43 06  &  1.4653   &  22.61  &  20.90  &  -11.43 \\
 144  &            &  dE      &  3  &        &  13 55 35.6  &   38 47 44  &  1.5525   &  21.96  &  21.24  &  -11.13 \\
 145  &  UGC 8742  &  Sd      &  0  &  2263  &  13 49 37.5  &   38 55 15  &  1.5535   &  19.40  &  15.15  &  -17.23 \\
 146  &            &  dE/I    &  2  &        &  14 00 48.6  &   40 58 30  &  1.5612   &  22.67  &  20.50  &  -11.85 \\
 147  &  SDSS J13  &  dE/I    &  0  &  2417  &  13 49 00.0  &   41 37 58  &  1.5854   &  18.23  &  17.00  &  -15.34 \\
 148  &  UGC 8793  &  Sd      &  0  &  2433  &  13 52 35.1  &   38 42 20  &  1.5959   &  18.03  &  14.53  &  -17.84 \\
 149  &            &  dI      &  3  &        &  13 57 30.2  &   38 52 00  &  1.6227   &  21.41  &  18.70  &  -13.67 \\
 150  &  UGC 8877  &  Sdm     &  0  &  2379  &  13 57 07.0  &   41 47 32  &  1.6546   &  19.05  &  14.21  &  -18.12 \\
 151  &  Mrk 1484  &  S0/a    &  0  &  2695  &  13 48 45.1  &   41 42 45  &  1.6781   &  16.39  &  14.01  &  -18.33 \\
 152  &            &  dE/I    &  2  &        &  14 00 21.7  &   39 14 31  &  1.6865   &  22.52  &  20.07  &  -12.27 \\
 153  &            &  dI      &  2  &        &  14 00 21.7  &   39 14 21  &  1.6882   &  22.16  &  19.93  &  -12.42 \\
 154  &  NGC 5383  &  Sb      &  0  &  2270  &  13 57 04.8  &   41 50 47  &  1.7010   &  14.97  &  10.95  &  -21.38 \\
 155  &            &  dI      &  3  &        &  13 53 29.3  &   38 33 41  &  1.7316   &  20.08  &  17.05  &  -15.32 \\
 156  &            &  dE      &  3  &        &  13 57 02.4  &   41 54 57  &  1.7617   &  21.02  &  19.47  &  -12.86 \\
 157  &            &  Sd      &  0  &  2209  &  13 50 15.9  &   42 01 37  &  1.8369   &  18.76  &  16.30  &  -16.03 \\
 158  &            &  dI/VLS  &  2  &        &  14 01 21.7  &   39 05 10  &  1.9327   &  23.69  &  19.83  &  -12.52 \\
 159  &            &  E/dE    &  3  &        &  13 57 30.8  &   42 08 55  &  2.0112   &  20.12  &  18.18  &  -14.15 \\
 160  &            &  S0pec   &  3  &        &  13 51 22.7  &   38 18 37  &  2.0215   &  18.27  &  15.76  &  -16.60 \\
 161  &            &  dE      &  3  &        &  14 00 11.2  &   38 43 12  &  2.0315   &  21.54  &  20.04  &  -12.30 \\
 162  &            &  dE,N    &  3  &        &  13 52 57.2  &   38 15 01  &  2.0449   &  21.82  &  21.01  &  -11.35 \\
 163  &            &  dI      &  3  &        &  13 53 21.0  &   38 14 07  &  2.0578   &  21.79  &  19.98  &  -12.39 \\
 164  &            &  dE      &  3  &        &  13 50 09.9  &   38 17 48  &  2.0920   &  22.27  &  21.79  &  -10.57 \\
 165  &  NGC 5325  &  Sc      &  3  &        &  13 50 49.5  &   38 14 07  &  2.1175   &  18.60  &  14.91  &  -17.45 \\
 166  &            &  dI      &  3  &        &  13 50 43.7  &   38 13 19  &  2.1349   &  20.33  &  18.38  &  -13.98 \\
 167  &            &  dE      &  3  &        &  13 50 40.6  &   38 13 02  &  2.1419   &  20.34  &  18.87  &  -13.49 \\
 168  &            &  dI      &  3  &        &  13 52 17.5  &   38 09 23  &  2.1479   &  19.90  &  17.51  &  -14.85 \\
 169  &            &  dI      &  3  &        &  13 54 48.0  &   42 25 54  &  2.1542   &  22.96  &  21.67  &  -10.66 \\
 170  &            &  dE/I    &  3  &        &  14 02 29.4  &   41 36 40  &  2.1707   &  22.66  &  21.48  &  -10.87 \\
 171  &            &  dI      &  3  &        &  13 59 39.8  &   38 27 44  &  2.1812   &  21.51  &  17.80  &  -14.54 \\
 172  &            &  Scd     &  0  &  2632  &  14 00 24.3  &   38 31 10  &  2.2151   &  16.88  &  14.28  &  -18.07 \\
 173  &    KUG 1358+387&  Sm      &  0  &  2609  &  14 00 25.3  &   38 31 15  &  2.2159   &  16.88  &  14.24  &  -18.11 \\
 174  &            &  dE      &  3  &        &  14 02 07.8  &   38 48 46  &  2.2212   &  21.72  &  20.14  &  -12.20 \\
 175  &            &  dI      &  3  &        &  14 02 18.0  &   38 38 31  &  2.3616   &  22.35  &  20.17  &  -12.17 \\
 176  &            &  dE,N    &  3  &        &  13 55 41.5  &   37 58 05  &  2.3641   &  22.31  &  21.32  &  -11.03 \\
 177  &            &  dE      &  3  &        &  13 53 06.7  &   37 54 11  &  2.3908   &  21.25  &  19.98  &  -12.39 \\
 178  &  UGC 8919  &  S0/a    &  0  &  2732  &  13 59 57.1  &   38 12 03  &  2.4324   &  15.93  &  13.60  &  -18.73 \\
 179  &            &  dE      &  3  &        &  13 51 40.0  &   37 53 02  &  2.4329   &  21.57  &  20.43  &  -11.93 \\
 180  &  NGC 5403  &  Sb      &  0  &  2746  &  13 59 51.1  &   38 10 58  &  2.4383   &  16.45  &  11.80  &  -20.53 \\
 181  &            &  dE      &  2  &        &  13 53 31.6  &   37 50 06  &  2.4580   &  20.59  &  17.98  &  -14.38 \\
 182  &            &  dI      &  3  &        &  13 59 59.1  &   38 09 58  &  2.4655   &  23.08  &  19.81  &  -12.52 \\
 183  &  HS 1347+3811  &  dE/I    &  0  &  2828  &  13 49 12.3  &   37 56 44  &  2.4828   &  18.74  &  16.56  &  -15.78 \\
 184  &            &  dE      &  3  &        &  13 53 14.5  &   37 43 35  &  2.5669   &  22.76  &  21.74  &  -10.62 \\
 185  &            &  dI      &  3  &        &  13 56 19.2  &   37 42 14  &  2.6465   &  19.93  &  16.53  &  -15.83 \\
 186  &            &  dI      &  3  &        &  13 56 29.5  &   37 42 13  &  2.6538   &  20.27  &  17.70  &  -14.64 \\
 187  &            &  dI      &  3  &        &  13 52 11.4  &   37 38 15  &  2.6663   &  20.88  &  19.19  &  -13.18 \\
 188  &            &  dE      &  3  &        &  13 47 55.5  &   37 45 26  &  2.7456   &  22.42  &  20.84  &  -11.50 \\
 189  &  IC 4341   &  Sc      &  0  &  2341  &  13 53 34.2  &   37 31 21  &  2.7706   &  17.07  &  14.31  &  -18.04 \\
 190  &            &  dI      &  3  &        &  13 52 47.9  &   37 30 21  &  2.7899   &  21.73  &  19.66  &  -12.70 \\
 191  &  UGC 8795  &  Sd      &  0  &  2302  &  13 52 48.5  &   37 29 27  &  2.8048   &  17.48  &  14.00  &  -18.36 \\
 192  &            &  dE      &  3  &        &  13 56 57.0  &   37 34 04  &  2.8060   &  21.02  &  19.18  &  -13.15 \\
 193  &            &  dE      &  3  &        &  13 53 23.5  &   37 27 56  &  2.8275   &  22.30  &  21.49  &  -10.86 \\
 194  &            &  dE      &  3  &        &  13 55 01.2  &   37 28 15  &  2.8381   &  22.37  &  21.35  &  -11.00 \\
 195  &            &  dE,N    &  3  &        &  13 58 06.0  &   37 34 22  &  2.8614   &  22.11  &  21.02  &  -11.32 \\
 196  &            &  dE      &  3  &        &  13 54 07.4  &   37 24 32  &  2.8870   &  21.76  &  20.47  &  -11.87 \\
 197  &            &  E/dE    &  3  &        &  13 58 13.0  &   37 30 45  &  2.9256   &  20.99  &  18.84  &  -13.50 \\
 198  &            &  dE      &  3  &        &  13 58 43.1  &   37 30 48  &  2.9559   &  20.29  &  17.97  &  -14.37 \\
 199  &            &  dI      &  3  &        &  13 58 43.0  &   37 30 17  &  2.9639   &  20.46  &  17.98  &  -14.36 \\
 200  &            &  dI      &  3  &        &  13 59 57.0  &   37 34 02  &  2.9947   &  21.04  &  18.92  &  -13.41 \\
 201  &            &  dE      &  3  &        &  13 54 02.9  &   37 17 57  &  2.9960   &  21.70  &  20.34  &  -12.00 \\
 202  &            &  dI      &  3  &        &  13 58 55.5  &   37 27 07  &  3.0271   &  20.28  &  18.36  &  -13.98 \\
 203  &            &  dE      &  3  &        &  13 59 17.9  &   37 28 43  &  3.0276   &  22.32  &  21.22  &  -11.12 \\
 204  &            &  dI      &  3  &        &  13 59 13.2  &   37 28 15  &  3.0294   &  19.89  &  17.82  &  -14.52 \\
 205  &            &  dE,N    &  3  &        &  13 56 23.1  &   37 18 39  &  3.0344   &  21.49  &  20.88  &  -11.46 \\
 206  &            &  dE,N    &  3  &        &  13 59 38.0  &   37 27 01  &  3.0779   &  22.69  &  22.22  &   -10.11 \\
 207  &            &  Sm      &  3  &        &  13 59 38.0  &   37 26 35  &  3.0846   &  19.82  &  15.96  &  -16.37 \\
 208  &            &  dE      &  3  &        &  13 57 52.6  &   37 17 24  &  3.1196   &  20.06  &  18.27  &  -14.07 \\
 209  &            &  dI      &  3  &        &  13 55 51.3  &   37 00 55  &  3.3098   &  22.54  &  20.84  &  -11.49 \\
 210  &            &  dE/I    &  3  &        &  13 57 30.9  &   37 02 09  &  3.3483   &  20.13  &  18.34  &  -14.00 \\
 211  &            &  dI      &  3  &        &  14 00 16.6  &   37 02 28  &  3.5031   &  21.03  &  19.57  &  -12.76 \\
 212  &            &  dI      &  3  &        &  13 55 37.0  &   36 43 18  &  3.5953   &  21.55  &  20.23  &  -12.12 \\
 213  &            &  dE      &  2  &        &  14 01 57.9  &   37 01 55  &  3.6433   &  21.90  &  18.83  &  -13.51 \\
 214  &  Mrk 465   &  S0/a    &  0  &  2700  &  14 01 24.0  &   36 48 00  &  3.8082   &  15.04  &  13.43  &  -18.90 \\
 215  &            &  dI      &  3  &        &  14 02 20.0  &   36 40 53  &  3.9895   &  21.23  &  19.65  &  -12.68 \\
\hline
\enddata
\end{deluxetable}

\begin{deluxetable}{llll}
\tablecaption{Evidence of Recent Star Formation in SDSS Spectra}
\label{spectra}
\tablewidth{0in}
\tablehead{\colhead{N5353--} & \colhead{$M_R$} & \colhead{Type} & \colhead{NED Name}}
\startdata
\multicolumn{4}{l}{Balmer series H$\alpha$ through H$\delta$ in emission} \\
\hline
019 & $-18.51$ & S0/a & Mrk 462 \\
026 & $-17.59$ & dE/I  & 2MASX J135340 \\
004 & $-21.35$ & Sb    & NGC 5350 \\
038 & $-16.44$ & dE/I  &  \\
030 & $-17.39$ & Sm   & KUG 1350+399 \\
050 & $-15.50$ & dE/I  &  \\
040 & $-16.35$ & Im     &  \\
014 & $-19.24$ & Sb    & IC 4336 \\
023 & $-17.91$ & Sdm & UGC 8721 \\
046 & $-15.76$ & dE/I  &  \\
051 & $-15.42$ & Sm   & LEDA 166181 \\
011 & $-19.34$ & Sc    & UGC 8736 \\
029 & $-17.39$ & S0/a &  \\
\hline
\multicolumn{4}{l}{H$\delta$, H$\gamma$ pass into absorption} \\
\hline
025 & $-17.65$ & Sm   & UGC 8807 \\
028 & $-17.40$ & S0/a & KUG 1347+399  \\
039 & $-16.43$ & Sm   &  \\
058 & $-14.82$ & dE/I  &  \\
\hline
\multicolumn{4}{l}{H$\beta$ passes into absorption} \\
\hline
034 & $-16.85$ & dE/I  &  \\
018 & $-18.56$ & Sc    & UGC 8841 \\
001 & $-22.30$ & Sbc  & NGC 5371 \\
005 & $-21.26$ & S0/a & NGC 5311 \\
021 & $-18.33$ & Sa    & KUG 1351+402 \\
027 & $-17.41 $ & S0/a & NPM1G +40.0334 \\
048 & $-15.59 $ & Sm   &  \\
007 & $-20.90 $ & Sc    & NGC 5313  \\
043 & $-15.99 $ & dE/I  &  \\
008 & $-20.35 $ & Sc    & NGC 5320 \\
041 & $-16.15$ & dE/I  &  \\
033 & $-16.96$ & S0/a &  \\
\hline
\multicolumn{4}{l}{Balmer series in absorption; [SII], N[II] emission} \\
\hline
012 & $-19.26$ & E      & NGC 5355 \\
045 & $-15.82$ & dE/I  &  \\
057 & $-14.94$ & Im    &  \\
\hline
\multicolumn{4}{l}{Strong H$\gamma,\delta$ absorption; hint of [SII], [NII] emission} \\
\hline
015 & $-19.17$ & Scd  & NGC 5346 \\
017 & $-18.63$ & E      & CGCG 218-060 \\
\hline
\multicolumn{4}{l}{Pop II spectra} \\
\hline
016 & $-18.99$ & S0    & NGC 5358 \\
031 & $-17.36$ & E      &  \\
032 & $-17.22$ & dE    &  \\
036 & $-16.49$ & E      &  \\
042 & $-16.15$ & E      &  \\
052 & $-15.20$ & dE    &  \\
053 & $-15.07$ & dE    &  \\
059 & $-14.69$ & dE    &  \\
\hline
\multicolumn{4}{l}{No SDSS spectra} \\
\hline
002 & $-21.77$ & S0    & NGC 5353 \\
003 & $-21.70$ & S0    & NGC 5354 \\
006 & $-20.97$ & Sa    & NGC 5326 \\
009 & $-20.16$ & Sc    & NGC 5362 \\
010 & $-19.93$ & Sab  & NGC 5337 \\
013 & $-19.25$ & Sd    & UGC 8726 \\
020 & $-18.47$ & Sdm & SDSS J134750 \\
022 & $-18.28$ & dE    & UGC 8840 \\
024 & $-17.91$ & S0    & CGCG 219-021 \\
037 & $-16.47$ & E      & LEDA 099754 \\
047 & $-15.68$ & Scd  &  \\
\enddata
\end{deluxetable}%

\begin{deluxetable}{lccc}
\tablecaption{Properties of 3 Groups}
\label{groups}
\tablewidth{0in}
\tablehead{   & \colhead{NGC 1407} & \colhead{NGC 5846} & \colhead{NGC 5353/4}}
\startdata
Nearby Galaxies Catalog designation                 &  51--8 & 41--1 & 42--1 \\
Distance (Mpc)                                                          &   25     &   26   &   29     \\
No. ${\rm T} \leq 1$~$M_R<-19$ (early; bright)  &  13      &   11   &    5      \\ 
No. ${\rm T} > 1$~$M_R<-19$  (late; bright)        &   1       &     4   &    10      \\
$\sigma_V$ (\kms)                                                    & $387 \pm 65$ & $320 \pm 35$ & $205 \pm 28$  \\
$2^{nd}$ turnaround $r_{2t}$ (kpc)                       & 900  & 840  & 530    \\
$L_R$ ($10^{11} \Lsun$)                                        &   2.1     &  2.6   &  2.0    \\
$M_T$ ($10^{13} \Msun$)                                       &  $7.3 \pm 2.7$ &  $8.4 \pm 2.0$ &  $2.1 \pm 0.5$  \\
$M/L_R$ ($\ML$)                                                      & $340 \pm 130$ & $320 \pm 80$ & $105 \pm 35$  \\
No.~Mpc$^{-2}$~ $M_R < -17$ at 200 kpc          & 34 &  29 &   42 \\
No. members                                                              & 250    & 251   &  137  \\
dwarf/giant ratio  ($>-17 / <-17$)                            & $6.5 \pm 1.3$ & $7.3 \pm 0.7$ & $2.9 \pm 0.6$ \\
Faint end slope $\alpha$                                         & $-1.43 \pm 0.05$&$-1.34 \pm 0.08$&$-1.15 \pm 0.03$ \\
Cutoff $M_R^{\ast}$                                                  &   $-$   & $-24.0$  &  $-21.9$ \\
Brightest galaxy                                                         & $-23.16$ & $-22.49$ & $-22.30$ \\  
\enddata
\end{deluxetable}%

\bibliographystyle{apj}
\end{document}